\documentclass[a4paper,fleqn,usenatbib]{mnras}

\usepackage{newtxtext,newtxmath}
\usepackage{lineno}
\usepackage[T1]{fontenc}
\usepackage{ae,aecompl}
\usepackage{xspace}

\usepackage{graphicx}	% Including figure files
\usepackage{amsmath}	% Advanced maths commands
\usepackage{amssymb}	% Extra maths symbols
\usepackage{ltxtable}

\newcommand{\gray}{${\gamma}$-ray}

\def \gRay{$\gamma$-ray\xspace}
\def \gRays{$\gamma$-rays\xspace}
\def \Fermi {{\it Fermi}\xspace}

\newcommand{\be}{\begin{equation}}
\newcommand{\ee}{\end{equation}}
\newcommand{\ba}{\begin{eqnarray}}
\newcommand{\ea}{\end{eqnarray}}

\title[A candidate list of high-energy-cosmic-ray accelerating blazars]{
Gamma-ray counterparts of 2WHSP high-synchrotron-peaked BL~Lac objects 
as possible signatures of ultra-high-energy cosmic-ray emission
}

\author[M. W. Toomey et al.]{
Michael W. Toomey,$^{1,2,3,4}$\thanks{E-mail: michael\_toomey@brown.edu}
Foteini Oikonomou,$^{5,6,7,3}$\thanks{E-mail: foteini.oikonomou@ntnu.no}
Kohta Murase$^{3,4,8}$\thanks{E-mail:murase@psu.edu}
\\
$^{1}$Brown Theoretical Physics Center, Brown University, Providence, RI 02912, USA\\
$^{2}$Brown University, Department of Physics, Providence, RI 02912, USA\\
$^{3}$The Pennsylvania State University, Department of Physics, University Park, PA 16802, USA\\
$^{4}$The Pennsylvania State University, Department of Astronomy \& Astrophysics, University Park, PA 16802, USA\\
$^{5}$Technische Universit{\"a}t M{\"u}nchen, Physik-Department, 
James-Frank-Str. 1, D-85748 Garching bei M{\"u}nchen, Germany\\
$^{6}$European Southern Observatory, Karl-Schwarzschild-Str. 2, Garching bei Munchen D-85748, Germany\\
$^{7}$Institutt for fysikk, NTNU, Trondheim, Norway\\
$^{8}$Yukawa Institute for Theoretical Physics, Kyoto University, Kyoto 606-8502, Japan
}

\date{Accepted XXX. Received YYY; in original form ZZZ}

\pubyear{2020}

\begin{document}
\label{firstpage}
\pagerange{\pageref{firstpage}--\pageref{lastpage}}
\maketitle

\begin{abstract}
We present a search for high-energy \gRay emission from 566 Active
Galactic Nuclei at redshift $z > 0.2$, from the 2WHSP catalog
of high-synchrotron peaked BL~Lac objects with eight years of
\Fermi-LAT data. We focus on a redshift range
where electromagnetic cascade emission induced by ultra-high-energy
cosmic rays can be distinguished from leptonic emission based on the
spectral properties of the sources.  Our analysis leads to the
detection of 160 sources above $\approx ~ 5\sigma$ (TS $\geq 25$) in
the 1 -- 300 GeV energy range.  By discriminating significant sources
based on their \gray{} fluxes, variability properties, and photon index
in the {\it Fermi}-LAT energy range, and modeling the expected
hadronic signal in the TeV regime, we select a list of promising
sources as potential candidate ultra-high-energy cosmic-ray
emitters for follow-up observations by Imaging Atmospheric Cherenkov
Telescopes.
\end{abstract}


%%%%%%%%%%%%%%%%%%%%%%%%%%%%%%%%%%%%%%%%%%%%%%%%%%

%%%%%%%%%%%%%%%%% BODY OF PAPER %%%%%%%%%%%%%%%%%%

\section{Introduction}
The {\it Fermi} Gamma-ray Space Telescope~\citep{2009ApJ...697.1071A}
and Imaging Atmospheric Cherenkov Telescopes (IACTs) have dramatically
increased the number of known \gRay sources, as well as our knowledge
of the non-thermal Universe. Among the observed extragalactic \gRay
sources, blazars, active galactic nuclei (AGNs) with jets aligned with
the observer's line of sight~\citep[e.g.,][]{Urry:1995mg}, are by far
the most numerous. They exhibit superluminal motion, and are some of the most powerful steady
sources in the Universe. Additionally, they dominate the \gRay sky and play an
important role in the energy budget of the
Universe~\citep{2019PhRvD..99f3012M}.

Categorized as either BL~Lacertae (BL~Lac) objects or Flat Spectrum
Radio Quasars (FSRQs), based on the properties of their optical
spectra, blazars are among the brightest objects in the
Universe. Blazars possess spectral energy distributions (SEDs) with a characteristic
double hump shape. 
The lower energy peak is generally thought to be powered by the synchrotron
emission of electrons in the blazar jet. The origin of the high-energy peak is a subject debate. In conventional leptonic scenarios, the \gRay{} emission is assumed to be powered by
inverse Compton radiation~\citep{Maraschi:1992iz,Sikora:1994zb}, but it could also have a hadronic origin~\citep{2000NewA....5..377A,MP01}.
BL~Lac objects are further sub-classified according to the
value of the frequency (in the source rest-frame) at which the
synchrotron peak of the SED occurs.
Low-energy ($\nu_S < 10^{14}$~Hz), intermediate-energy ($10^{14} <
\nu_S < 10^{15}$~Hz) and high-energy ($\nu_S > 10^{15}$~Hz)
synchrotron peaked, referred to in short as LSP, ISP and HSP
respectively~\citep{Padovani:1994sh,2010ApJ...716...30A}. 

More recently, observations with IACTs have revealed an additional class of
BL~Lac objects, whose spectrum in the \Fermi-LAT energy range is
hard (meaning that the spectral index in this energy range $\gamma <
2$), placing their peak of the high-energy ``hump'' in the SED, after
accounting for absorption during their extragalactic propagation, in
the TeV energy range. Additionally, these sources typically possess
$\nu_S > 10^{17}$~Hz. These properties are suggestive of extreme
particle acceleration which has led to them being referred to as
extreme HSPs~\citep{2001A&A...371..512C,2018MNRAS.477.4257C,2020NatAs...4..124B}.

One of the greatest mysteries in particle astrophysics today is the
origin of ultra-high-energy cosmic-rays (UHECRs). Such particles are
observed with energies in excess of $\approx$ $10^{20}$ eV~\citep[see e.g., reviews by][]{AlvesBatista:2019tlv,Anchordoqui:2018qom,Mollerach:2017idb,Kotera:2011cp}. Since the Larmor radius for cosmic-rays above the \textit{ankle}, at $\sim 4\times 10^{18}$~eV, exceeds that of the Galaxy, these UHECRs are very likely of extragalactic origin. One of the most promising candidates proposed as the sources of UHECRs are AGNs~\citep{1964ocr..book.....G,Hillas:1985is}. Blazars, specifically, have often been proposed as sites of UHECR
acceleration~\citep[see e.g.,][and references
  therein]{Dermer:2010iz,Murase:2011cy,Rodrigues:2017fmu}.

This is particularly true for extreme HSP BL Lac objects.  Attempts to describe the origin of their hard TeV spectra
with standard leptonic models often require extreme
parameters~\citep[e.g.,][]{Katarzynski:2006bd}. An
alternative scenario for the observed \gRay emission of extreme HSPs is that it is secondary emission from UHECRs
~\citep[e.g.,][]{2010APh....33...81E,Essey:2010er,Murase:2011cy,Takami:2013gfa,Aharonian13,Oikonomou:2014xea,2016ApJ...817...59T,2019MNRAS.483.1802T}. If these sources produce UHECRs, these will interact with photons from the extragalactic background light (EBL) and the cosmic
microwave background (CMB), and produce electron-positron pairs (Bethe-Heitler) emission, and pionic $\gamma$-rays. However, the spectrum of UHECR-induced secondary \gRays is expected to extend to higher energies than in standard leptonic scenarios due to the continuous injection of high-energy leptons via the Bethe-Heitler (BH) pair-production process during intergalactic propagation, leading to a natural explanation for the observed hard TeV spectra. 

Primary and secondary \gRays with energy $E_{\gamma}\sim (m_e^2
c^4)/ \varepsilon_{\gamma} \sim 260 ~ (1\,{\rm eV}/\varepsilon_{\gamma})$~GeV, with photons of the EBL and the CMB with energy $\varepsilon_{\gamma}$, which results in the production of electron-positron
pairs~\citep{Gould:1966pza,Stecker:1992wi}. Electrons and positrons produced in interactions of \gRays with the EBL/CMB inverse Compton upscatter background photons, and generate secondary
\gRays. The two processes (pair-production and inverse Compton emission) produce an electromagnetic ``cascade'' in the intergalactic medium, until the
\gRays drop below the threshold for pair production on the EBL. During the intergalactic propagation of the cascade emission, the electrons (and positrons) get deflected in the presence of intergalactic magnetic fields (IGMFs), causing either magnetic broadening of the cascade emission, or, in the presence of a stronger field, a suppression of the observed GeV cascade flux from the source~\citep{Gould78,Aharonian:1993vz}. The detection (or absence of) the cascade emission from TeV emitting blazars can result in the measurement of (or lower bound on) the IGMF in the line of sight from these sources.  Similarly the detection of magnetic broadening, ``halo'' emission, can result in a measurement of the strength of IGMFs~\citep[e.g.,][]{NS09,Elyiv09,2014A&A...562A.145H,2017ApJ...835..288A}. High-redshift, hard-spectrum blazars, whose intrinsic spectrum extends beyond TeV energies are thus ideal sources to look for the signatures of the effects of the IGMF (cascade flux, and halo component), as the cascade emission in the GeV spectrum dominates the observed emission in this energy band. This combination of parameters maximises the expected cascade emission in these sources, and in the case of non-observation of the expected cascade emission, can result in lower limits on the IGMF strength. For the same reason, the expected halo component is maximal for high-redshift, hard-spectrum sources (although at the disadvantage of having an overall fainter source). The highest redshift blazars that have been detected to date with IACTs are PKS~1441+2~\citep{Abeysekara_2015} and S3~0218+35~\citep{Ahnen_2016}. These are of the FSRQ type, and were detected up to $\sim200$~GeV energy. They are thus not expected to exhibit the signature of UHECR emission which has been proposed for extreme HSPs in their \gRay spectra, but demonstrate the possible reach of future, deeper, IACT observations.

\subsection{Time Variability}

An important observational distinction between leptonic and UHECR-induced intergalactic cascade emissions is related to the different deflections experienced by the leptonic and UHECR beam in the IGMF. As a result, one expects different variability properties of the \gRay emission in these two cases. In the case of a leptonic beam, time delays are only relevant for the secondary (cascade) emission, and come from the deflections of the electrons produced in the interactions of the primary \gRays with background photons. Electrons with energy $E'_{\rm e}$ (in the cosmic rest frame at redshift $z$), and Lorentz factor $\gamma'_e = E'_e/(m_e c^2)$, traversing a region with magnetic field strength $B$, have a Larmor radius $r_{\rm L} \approx eB/E'_e$, where $e$ is the charge of the electron. Such electrons will experience deflections of order $\theta_{\rm IC} \approx \sqrt{2/3}(\lambda_{\rm IC}/r_{\rm L})$ before inverse Compton upscattering CMB photons, after, on average, one inverse Compton cooling length $\lambda_{\rm IC} = 3m_e c^2/(4\sigma_T U_{\rm CMB} \gamma'_e) \approx 70~{\rm kpc} ~(E'_{e}/5~{\rm TeV})^{-1}(1+z)^{-4}$, where $\sigma_{\rm T}$ is the Thompson cross-section and $U_{\rm CMB} \simeq0.26~{\rm eV}~(1+z)^4$ is the energy density of the CMB at redshift~$z$.

The typical time delay experienced by cascade photons due to the deflections of the electrons is of order,~\citep[e.g.,][]{Murase08b,Takahashi08,Dermer11}, $\Delta t _{\rm IC, \rm IGMF} / (1+z) \approx \theta_{\rm IC}^2 (\lambda_{\rm IC}+\lambda_{\rm \gamma \gamma})/2c$. Here, $\lambda_{\gamma \gamma}~\sim~20~{\rm Mpc}~(n_{\rm EBL}/0.1~{\rm cm^{3}})^{-1}$ is the average distance travelled by a primary \gRay before it interacts with an EBL photon to create an electron-positron pair. In the above expression, $n_{\rm EBL}$ is the number density of EBL photons, normalised to the value relevant for interactions with 10~TeV
\gRays. The electrons subsequently upscatter CMB photons via the inverse-Compton process, typically to energy $E_{\gamma} \approx(4/3) {\gamma}_e^{\prime 2} \varepsilon_{\rm CMB}$, where $\varepsilon_{\rm CMB}\approx2.8k_BT_{\rm CMB,0}$ and $k_{B}$ the Boltzmann constant, and $T_{0,\rm CMB}$ the temperature of the CMB at $z=0$. For such \gRays with energy $E_{\gamma}$ from a source at $z \ll 1$, we obtain a characteristic time delay~\citep[e.g.,][]{Murase:2009pg}, \small{
\ba \frac{\Delta t_{\rm IC, \rm IGMF}}{(1+z)} &\approx&\frac{\theta_{\rm IC}^2}{2c} (\lambda_{\rm IC}+\lambda_{\rm \gamma\gamma})\nonumber\\ \,\,\,\,\,\,\,\,\,\,\,\,\,\,\,\,\,\,\,\, &\sim&4\times 10^5~{\rm yr}~\left(\frac{E_{\gamma}}{0.1~{\rm TeV}}\right)^{-2} \left( \frac{B}{10^{-14}~{\rm G}}\right)^2
\left( \frac{\lambda_{\gamma\gamma}}{20~{\rm Mpc}}\right). \,\,\,\,\,\,\,\,\,\,\,\,
\label{eq:DeltaT}
\ea } \normalsize Thus, any detectable variable emission in short timescales 
is likely primary in origin. An additional, slowly variable component may exist, due to the reprocessed emission, if the IGMF is sufficiently low ($B \lesssim 10^{-17}$ G).

In the UHECR-induced intergalactic cascade scenario, there is no
primary \gRay component.  The main energy loss channel for protons
with energy less than $ 10^{19}$ eV is through the Bethe-Heitler
pair-production process. The trajectory of the protons may be regarded
as a random walk through individual scattering centers of size, $l_c
\sim ~\mathcal{O}(\rm Mpc)$.  The Larmor radius of an UHECR with
energy $E_p$ is, $r_{{\rm L},p}~\sim~eB/E_p~\sim~800~{\rm
  Mpc}(B/10^{17}~{\rm G})(E_p/10^{17}~{\rm eV})^{-1}$.  After crossing
a scattering center of size $l_c$ the proton experiences a deflection
of order $\theta_p \approx \sqrt{2/3} l_c/r_{{\rm L},p}$ and a time
delay of order $\Delta t_{0}\sim(l_c/c) \theta_p^2$. Thus, the total
time delay experienced after travelling the characteristic energy loss
length of the Bethe-Heitler process, $\lambda_{\rm BH}$, is of order,
$\Delta t_p \sim (\lambda_{\rm BH}/l_c){\Delta t}_{0} \sim
(\lambda_{\rm BH}/c) \theta_p^2 \sim 20~{\rm yr}~(E_p/10^{17}~{\rm
  eV}) (B/10^{-14}~{\rm G})^{2} (\lambda_{\rm BH}/{\rm 1~Gpc})
(1+z)$~\citep{Murase:2011cy,Prosekin12}.  At every interaction the
proton produces electrons (and positrons) which cascade down to GeV
energies over a length scale of order $\lambda_{\rm IC}$.  Thus, in
this case, the relevant time delay is ${\rm max}(\Delta t_{\rm IC,IGMF},\Delta t_p)$, i.e. the maximum of the delays
experienced by the protons and those of the leptonic cascade emission
given by Equation~\ref{eq:DeltaT}.  As a result, we expect that the
sources whose emission is dominated by UHECR cascades should be non-variable or at most slowly variable.

The search for variability is a strong motivation for the present
work.  However, though the absence or weakness of the variability is
one of the crucial signatures of the UHECR-induced cascade scenario,
negative detection itself is not proof for the
UHECR origin of the emission.  This is because, even if the
variability is present, it may simply be below the experimental
sensitivity.
 
Our goal is to identify sources which have a hard spectrum in the GeV energy range. Such sources are good candidates for very high-energy (VHE) follow-up observations, which can reveal TeV spectra of extreme HSP blazars, and possibly the signatures of UHECR acceleration and information on IGMFs. Additionally, we investigate variability properties of our source sample. The presence of the variability can rule out hadronic origin of the \gRay emission. The nondetection of variability is harder to interpret in this context, as it can be caused by either intergalactic propagation effects or insufficient sensitivities of the instruments. Nevertheless, we make inferences by examining the entire sample, and trends as a function of redshift and \gRay flux.

The layout of the paper is as follows. In Section~2 we discuss the selection of our catalog and the analysis of \Fermi-LAT data, including the calculation of the time variability.  In Section~3 we discuss our modeling of the leptonic and hadronic emissions in the TeV regime.  In Section~4 we present the results of our work and in Section~5 we discuss our results and propose avenues for further study.

\section{Analysis}
\subsection{Data Selection}
For our analysis we opt to use the 2WHSP catalog of HSP blazars~\citep{Chang:2016mqv}, which was, until recently, the most complete catalogue of HSP and candidate extreme-HSP sources (see discussion in Section~\ref{sec:discussion}). All sources from the catalog with redshifts $\geq 0.2$ were selected and submitted to a full {\it Fermi} analysis, a total of 566 sources. This was the only cut made on the 2WHSP catalog, motivated by the theoretical prediction that UHECR-induced cascade emission can be more prominent for higher redshift HSPs~\citep{Murase:2011cy,Takami:2013gfa}. 
Note that some sources in the catalog only have a redshift with a lower limit. We still include these in the analysis but interpret results differently where relevant.

This sample of high redshift blazars from the 2WHSP catalog forms the basis of the sources studied in this analysis. An analysis searching for \gRay emission was conducted by the authors of the 1BIGB catalog~\citep{Arsioli:2016ewb} (see also~\citealp{Arsioli:2018yeq}). However, these analyses did not include a study of source variability. In order to determine potential cosmic-ray accelerators based on VHE \gRays, this information can be critical. In Section 4 we elaborate further on how our results compare to and differ from those of~\citet{Arsioli:2016ewb}. 

Many of the sources studied in this analysis have a counterpart in the 3FGL~\citep{Acero:2015hja} and the recently released 4FGL catalog~\citep{2019arXiv190210045T}. However, the variability analysis from the 4FGL, or the previous 3FGL, is not appropriate for of our work
due to the inclusion of sub-GeV photons. In the UHECR-induced cascade scenario, cascade emission is expected to be dominant in the 10--100 GeV range as we show in the following sections. At the sub-GeV energy range, because of the IGMF suppression of the cascade component, primary photons that are intrinsically variable would dominate the spectrum for blazars. However, we would not have sufficient statistics for the variability analysis if we focused on the high-energy data in the 30--100 GeV range only. Therefore in this work we have chosen the 1-100 GeV energy range for our analysis. This is not ideal but still useful to see the variability and discriminate between the leptonic and hadronic scenarios. We further note that since the sources of interest in our study have a hard spectrum in the {\it Fermi} energy range, it would be more challenging to detect the variability in the 100 MeV-1 GeV energy range than for the average of 4FGL sources.

\subsection{Data Analysis}
An unbinned maximum likelihood approach was used in this work for spectral analysis utilizing {\it Fermi} Science Tools \href{https://fermi.gsfc.nasa.gov/FTP/fermi/software/ScienceTools/v10r0p5/ReleaseNotes_v10r0p5.txt}{v10r0p5}. For each blazar candidate, a region of interest $8^\circ$ in radius was created from Pass 8 SOURCE class ($evclass=128$) photons that where detected on both the FRONT and BACK of the LAT detectors ($evtype=3$). Data were filtered temporally from 5 August 2008 (239587201 MET) to 24 September 2016 (496426332 MET) culminating in a total of 8.2 years of data. Data where additionally filtered by considering photons of energies 1.0 -- 300 GeV and setting a maximum zenith angle of $100^\circ$ to avoid atmospheric background. Periods where data taken from LAT were of poor quality where removed utilizing the tool \textit{gtmktime}. At this step, the LAT team's recommended filter
expression for SOURCE class photons
(DATA\_QUAL$>0$)\&\&(LAT\_CONFIG==1) was used.

For each source an exposure hypercube was calculated -- a measure of the amount of time a position on the sky has spent at a certain inclination angle. The exposure hypercube was computed with \textit{gtltcube} by binning the off-axis angle in increments of 0.025 cos$\theta_{OA}$ and setting the spatial grid size to $1^\circ$. For our analysis, we follow the {\it Fermi}-LAT recommendation to implement the zenith angle cut for exposure during the calculation of the exposure hypercube as opposed to during the determination of good time intervals with \textit{gtmktime}. The exposure map was then calculated for a region $18^\circ$ in radius, with 72 latitudinal and longitudinal points, and 24 logarithmically uniform energy bins.

We conducted a sanity check of our own implementation of {\it Fermi}
Tools used in this analysis by utilizing the Fermipy package
\citep{2017ICRC...35..824W}. The results obtained with the two methods
were found to be consistent. We did not use the Fermipy
package for our results as an unbinned analysis method was not
available.

\subsection{Modeling}
For each candidate \gray{} blazar, a model was constructed using known
3FGL sources, the Galactic diffuse emission model
\textit{gll\_iem\_v06}, and isotropic diffuse model
\textit{iso\_P8R2\_SOURCE\_V6\_v06} with
\textit{make3FGLxml}. Non--3FGL sources were added to the model as a
simple power law,

\begin{equation}
\frac{dN}{dE} = N_0\Bigg[\frac{E}{E_0}\Bigg]^{-\gamma}
\end{equation}

\noindent The photon index, $\gamma$, and normalizations, $N_0$, were
set free to be fit but the pivot energy, $E_0$, was fixed at 3.0 GeV
for non--3FGL sources and set to 3FGL catalog values otherwise. The
normalizations of the Galactic and isotropic diffuse emissions were
fit, in addition to the normalization of all sources within $3^\circ$
and variable sources out to $5^\circ$. The maximum likelihood for each
source was then computed using an unbinned technique with the
\textit{NEWMINUIT} minimizer implemented by the
\textit{UnbinnedAnalysis} module from {\it Fermi} Tools.

\subsection{Time Variability}
A detailed variability analysis was conducted for significant sources
- those above a $TS$ of 25. Variability was determined by analyzing the
sources at 60 day intervals with two different models. The first model
allows the source normalization to be optimized for each bin. In the
second model the source spectrum is fixed to correspond to the null
hypothesis, i.e. the source not being variable. For the first model,
if the flux in a temporal bin was not significant ($TS < 9$) or if
errors were larger than $\Delta F_i/F_i$, a 90\% confidence
Bayesian upper limit was calculated with the
\textit{IntegralUpperLimit} module. Our variability index corresponds
to that defined by {\it Fermi} in their 2FGL paper~\citep{2012ApJS..199...31N},
\begin{equation}
TS_V = 2 \sum_i \frac{\Delta F_i^2}{\Delta F_i^2 +
  f^2F^2_\text{c}}V_i^2,
\end{equation}
\noindent where $\Delta F_i$ is defined as the as the flux error, $F_c$ the flux
for the source if it was not variable, $V_i^2$ the difference in
log-likelihoods for the null and alternative hypothesis, and
\textit{f} is a systematic correction factor determined by the {\it Fermi} 
team to be 0.02 in this calculation. For bins with low $TS$ the
variability was calculated using a similar statistic,
\begin{equation}
TS_{\text{UL}} = 2 \sum_i \frac{0.5(F_{UL} - F_i)^2}{0.5(F_{UL} -
  F_i)^2 + f^2F^2_\text{c}}V_i^2.
\end{equation}
The variability index is distributed as a $\chi^2$ distribution
where the degree of freedom corresponds to the number of bins, here
50. Thus, a total $TS_V> 63.17$ implies less than 10\% chance
for the source to exhibit non--variability. For this analysis, a
source with an index above this value is considered to be variable.

\subsection{Source Selection Criteria}
In choosing promising sources for follow-up, a set of criteria were placed on the sources to establish merit. A primary cut was imposed on the flux of each source, $F~>~2.5~\times~10^{-10}$~cm$^{-2}$~s$^{-1}$, to eliminate dim sources. It is also important that the sources have a hard photon index, $\gamma < 1.8$, and have low variability, $TS_V < 70$. From these cuts a list of 12 potentially promising sources was compiled - see Table \ref{tab:example_table}. As a final means of selecting the most promising sources for observation, the hadronic and leptonic spectrum for each source was calculated and classified based on their detecability with IACTs. Sources which are detectable are put into two merit classes. Class I are likely detectable with current generation IACT detectors. Class II sources will likely take much longer for detection and are therefore better candidates for next generation detectors like CTA. Sources are marked as belonging to one of these two classes in Table \ref{tab:example_table}.

\section{TeV Spectrum Modeling}
\subsection{Leptonic Scenario}

With knowledge of the spectrum in the GeV regime, it is possible to
predict the expected spectrum at TeV energies. For leptonic emission
this can be done by assuming that the spectral index derived at GeV
energies and extending the maximum energy while
accounting for attenuation of $\gamma$-rays due to pair production on the EBL. In this scenario, the optical depth for EBL photons, $\tau_{\gamma \gamma}(E)$, is
dependent on the redshift of the source. Thus, we can model the
expected energy flux out to TeV energy, 
\begin{equation}
E F_E = {\mathcal N}_{\rm lep} \cdot E^{-\gamma_{\rm LAT}+2} \cdot e^{-\tau_{\gamma \gamma}(E)},
\end{equation}
where ${\mathcal N}_{\rm lep}$ and $\gamma_{\rm LAT}$ are the flux normalisation and spectral index in the LAT energy range as determined in our analysis. We have used data from \citet{Inoue12} to calculate the attenuation of primary leptonic \gray{}s.

\subsection{UHECR-Induced Cascade Scenario}
\begin{figure}
	\includegraphics[width=\columnwidth]{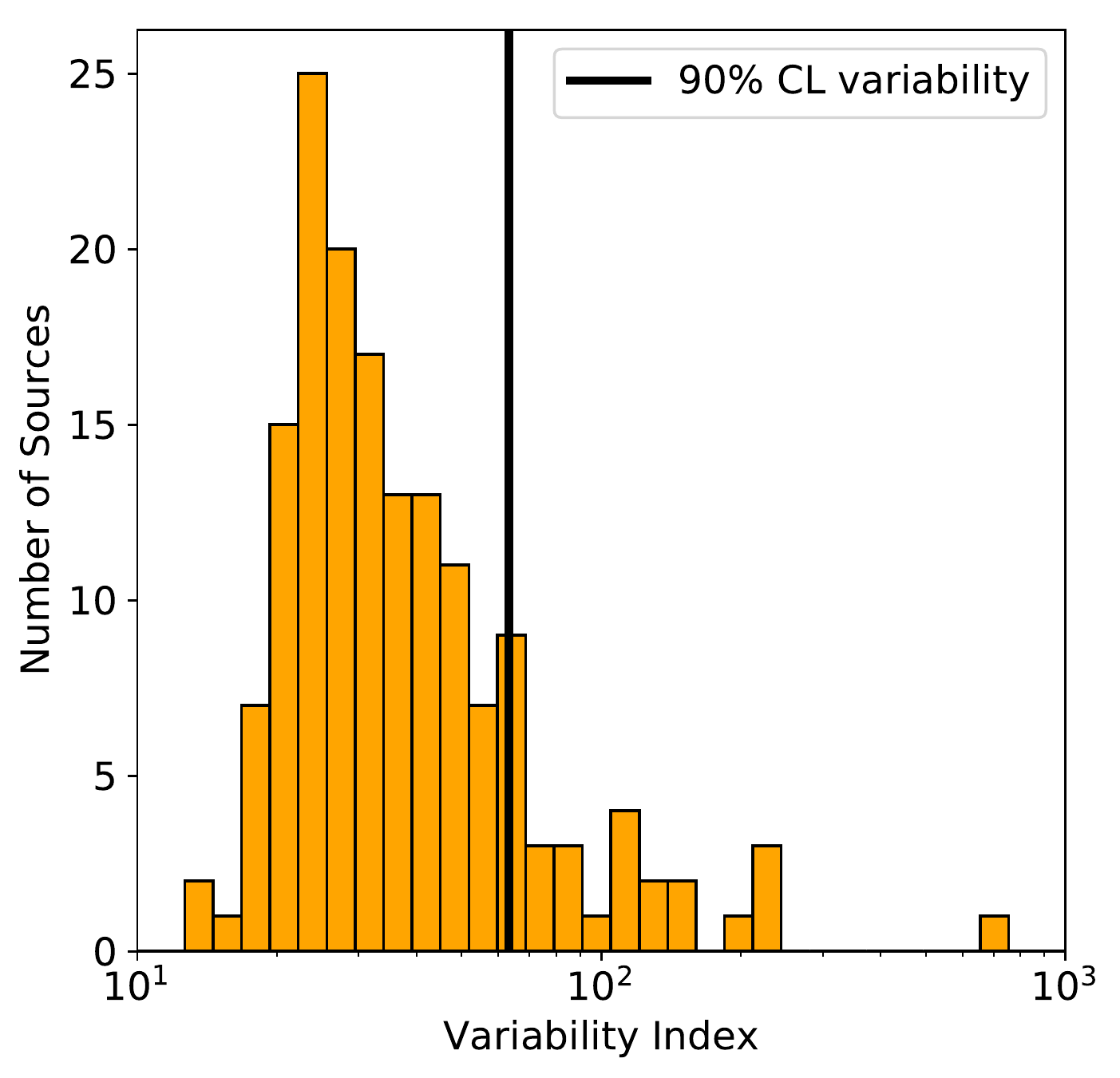}
    \caption{Distribution of the source variability index. The
      variability index for this analysis follows a $\chi^2$
      distribution with 50 degrees of freedom. Thus a source with a
      variability index in excess of 63.17 exhibits variability with
      90\% confidence.}
    \label{fig:var_dist}
\end{figure}
Similar to primary \gRays, cascades occur for UHECRs traveling
through intergalactic space. The observed TeV
spectrum, however, should be harder than in the leptonic case due to
the injection of high--energy leptons from the Bethe--Heitler
process. We adopt the analytic formula from \citep{MBT12} for such cascades~\citep[see also][]{Berezinsky:1975zz}. The
approximate spectrum for the cascade emission is given by,
\begin{equation}
E G_E \propto
\begin{cases}
(E/E_{\rm br})^{1/2} &(E \leq E_{\rm br}),
  \\ (E/E_{\rm br})^{2 - \beta} & (E_{\rm br} \leq E
  \leq E_{\rm cut}),
\end{cases}
\end{equation}
where the normalization is set by $\int dE G_E=1$, with $E_{\rm cut}$, the critical energy at which $\tau_{\gamma\gamma}(E)=1$ due to the EBL absorption,for pair-production on the EBL, 
$E_{\rm br}\approx4\varepsilon_{\rm CMB}{E}_e^{\prime 2}/(3 m_e^2)$ the energy below which the number of electrons remains constant, where $E'_e \approx (1+z)E_{\rm cut}/2$ and $\varepsilon_{\rm CMB}\approx2.8k_BT_{\rm CMB,0}$, and the cascade photon index $\beta\approx1.9$~\citep{MBT12}. 

The optical depth to BH pair-production, $\tau_{\rm BH}$ for cosmic rays around $10^{19}$~eV, is given by the approximate expression,
\begin{equation}
\tau_{\rm BH}\approx\frac{d}{\rm 1000~Mpc},
\end{equation}
where $d$ is the particle travel distance. 
Thus, the expected observed spectrum is given by,
\begin{equation}
E F_E = C_{\rm had} \cdot E G_E \frac{{\rm min}[1,\tau_{\rm BH}]}{\tau_{\gamma \gamma}(E)} \left( 1 -
\exp^{-\tau_{\gamma \gamma}(E)} \right),
\end{equation}
where $C_{\rm had}$ is the normalisation factor and this equation is implemented without a cutoff (because the cutoff shape is taken into account via $\tau_{\gamma\gamma}$). 
In addition, the low-energy spectrum is suppressed by IGMFs. The comparison with the point spread function of the {\it Fermi}-LAT~\citep[e.g.,][]{NS09,MBT12} suggests that the cascade emission is suppressed below $\sim30$ GeV for $B\sim 3 \times 10^{-17}$~G~\citep[see also equation 6 of][and discussion]{2011A&A...527A..54K}.

In practice, our procedure provides the shape of the UHECR-induced cascade spectrum but does not encode the expected differential energy flux. Thus, we normalize our hadronic cascade spectrum using the normalisation obtained through the \Fermi data in the $10-100$~GeV range.

\section{Results} 
\subsection{Likelihood and Variability Results}
\begin{table*}
    \scriptsize
	\centering
	\caption{Results from the analysis of 2WHSP sources.
          Included is the name from the 2WHSP catalog, whether it is a Class I or II source, the luminosity from 1 -- 300 GeV,
          $\boldmath L_{44}$, in units of $10^{44}$ erg s$^{-1}$, the 1 -- 300 GeV photon flux,
          $\boldmath {({\rm d}N/{\rm d}t)}_{-10}$, in units of $10^{-10}$ cm$^{-2}$ s$^{-1}$,
          the test statistic from the likelihood fit , $\boldmath{TS}$, the normalization
          and its error scaled by $10^{-11}$ cm$^{-2}$ s$^{-1}$ GeV$^{-1}$, $N_{-11}$ and $\sigma_{N}$,
          the photon index $\gamma$ and its error $\sigma_{\gamma}$, the variability index $TS_V$ (> 63.17 is a variable source),
          the source redshift $z$, alternative identifier, and the logarithm of the synchrotron peak frequency,
          $\log \nu_{\rm S, pk}$ in units of $\log{\text{Hz}}$. The latter three entries are obtained from the 2WHSP catalog.
          Listed here are the results based on our source selection criteria. The full table, with all sources from the analysis, is available in the Appendix.}
	\label{tab:example_table}
	\begin{tabular}{ccccccccccccc} % four columns, alignment for each
		\hline
		\textbf{2WHSP Name} & {\bf Class} & \boldmath$L_{44}$ & \boldmath${\left(\frac{dN}{dt}\right)}_{-10}$ & \boldmath$TS$ & \boldmath$N_{-11}$ & \boldmath$\sigma_{N}$ & \boldmath$\gamma$ & \boldmath$\sigma_{\gamma}$ & \boldmath${TS}_V$ & \textbf{z} &\textbf{Other Name} & \textbf{$\log \nu_{\rm S, pk}$} \\
		\hline
J011904.6-145858 & II    &    45.54    &    2.82    &    134.42    &    1.82    &    0.29    &    1.77    &    0.12    &    27.0    &    >  0.530  &   3FGL J0118.9–1457   &         16.1  \\
J050657.7-543503 & I &  25.07  &  5.05  &  488.2  &  2.21  &  0.21  &  1.56  &  0.07  &  40.6  &  >                0.260  & RBS 621 & 16.2 \\
J060408.5-481725  & II  &    20.5    &    2.69    &    132.76    &    0.26    &    0.05    &    1.73    &    0.11    &    37.1    &    >                0.370    &   1ES 0602-482   &         16.2  \\
J101244.2+422957 &   &    19.77    &    2.78    &    157.36    &    1.01    &    0.14    &    1.74    &    0.11    &    29.3    &                     0.365                     &   3FGL J1012.7+4229   &         16.8  \\
J103118.4+505335 & I &  41.89  &  9.22  &  1014.9  &  7.26  &  0.48  &  1.74  &  0.05  &  52.7  &  0.360  & 1ES 1028+511 & 17.0 \\
J112453.8+493409 & II    &    93.35    &    4.89    &    362.65    &    2.91    &    0.29    &    1.78    &    0.08    &    63.9    &    >                0.570                       &   RBS 981   &         16.5  \\
J124312.7+362743  & II &  99.06  &  21.98  &  2594.8  &  33.10  &  1.52  &  1.78  &  0.03  &  62.1  &  1.065   & Ton 116 & 16.2 \\
J141756.5+254324 & I  &  10.14  &  2.86  &  147.3  &  7.2  &  1.65  &  1.63  &  0.08  &  24.6  &                   0.237  & RBS 1366 & 17.4 \\
J143657.7+563924 &  II    &    57.3    &    5.79    &    522.39    &    8.28    &    0.8    &    1.77    &    0.06    &    47.4    &    >                0.430    &   RBS 1409  &         16.9  \\
J150340.6-154113 & I  &    65.09    &    9.18    &    439.69    &    5.04    &    0.42    &    1.79    &    0.07    &    52.0    &    >                0.380    &   RBS 1457   &         17.6  \\
J175615.9+552218  & II  &    45.88    &    3.68    &    223.86    &    3.85    &    0.5    &    1.76    &    0.08    &    32.0    &    >                0.470                      &   RGB J1756+553   &         17.3  \\
J205528.2-002116    &   &  32.19    &    2.96    &    92.1    &    0.8    &    0.14    &    1.75    &    0.13    &    27.7    &                     0.440    &                      3FGL J2055.2-0019   &         18.0  \\
		\hline
	\end{tabular}
\end{table*}

In our analysis of 566 VHE \gray{} blazar candidates above $z \geq
0.2$ from the 2WHSP catalog of HSP BL~Lac objects, we detected 160
sources above $\approx ~ 5\sigma$ ($TS\geq 25$). Our best fit
spectral parameters and {\it TS} values are in agreement with previous
analyses of these sources \citep{Arsioli:2016ewb,Acero:2015hja}.

Of the 160 \gRay detected sources, 26 were found to exhibit
variability with greater than 90\% confidence while 134 did not
present variability, see Figure \ref{fig:var_dist}. Based upon our
criterion, the majority of our sources do not exhibit significant
variability. Table \ref{tab:example_table} contains the results from
the analysis for the most promising sources identified. The entire table with
all sources from the analysis can be found in the Appendix.

Many of the sources in the 2WHSP catalog do not have firm redshifts. 
While some sources have precise measurements, some have only
lower limits, and others have measurements but uncertainties are still
large. Where relevant, we separate data based on the nature of the
redshift measurement.

\begin{figure}
\begin{center}
\includegraphics[width=\linewidth]{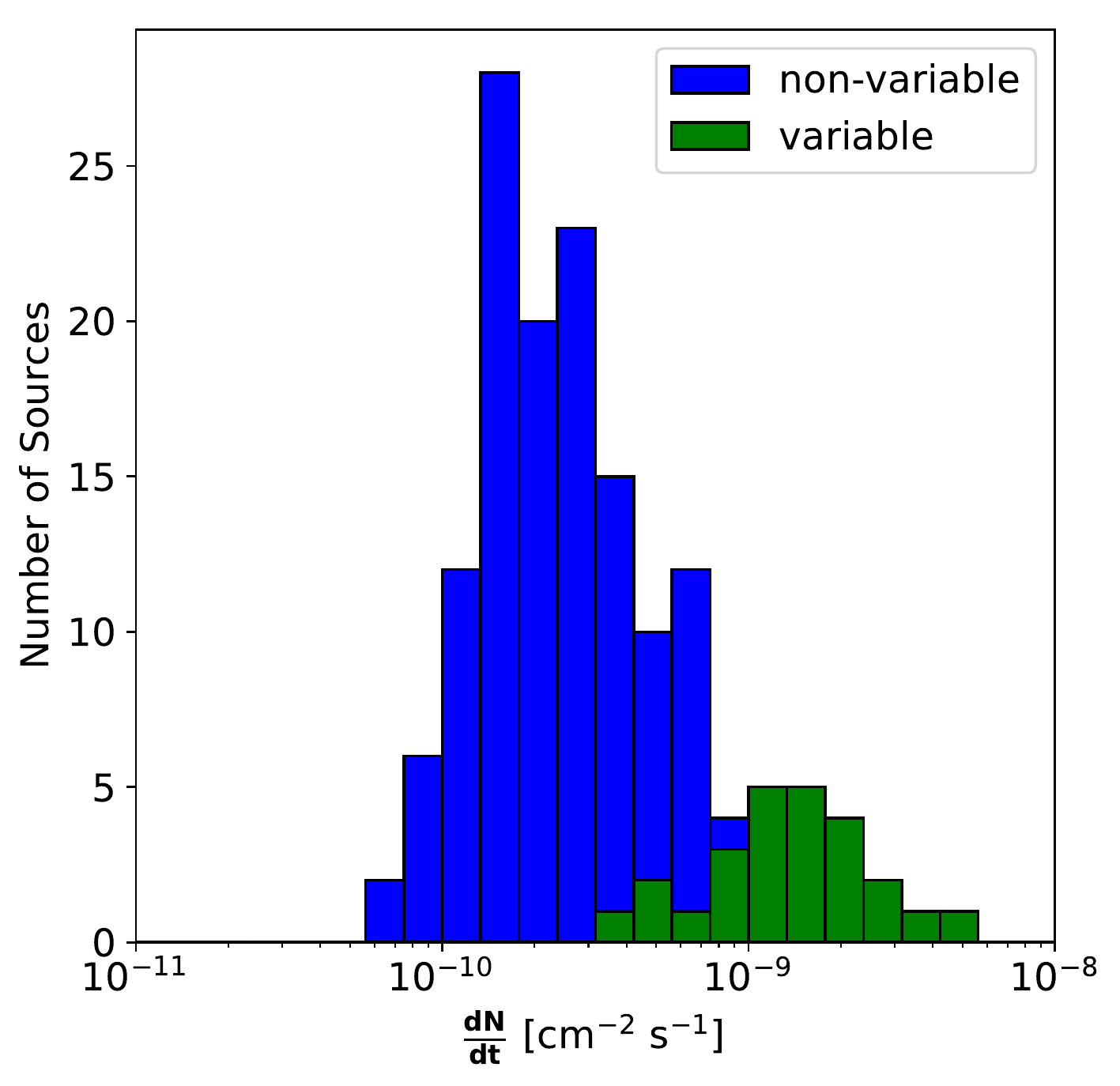}
\caption{Distribution of fluxes over the 1 - 300 GeV band for non--variable and variable sources.}
\label{fig:flux_dist}
\end{center}
\end{figure}

\begin{figure}
\begin{center}
\includegraphics[width=\linewidth]{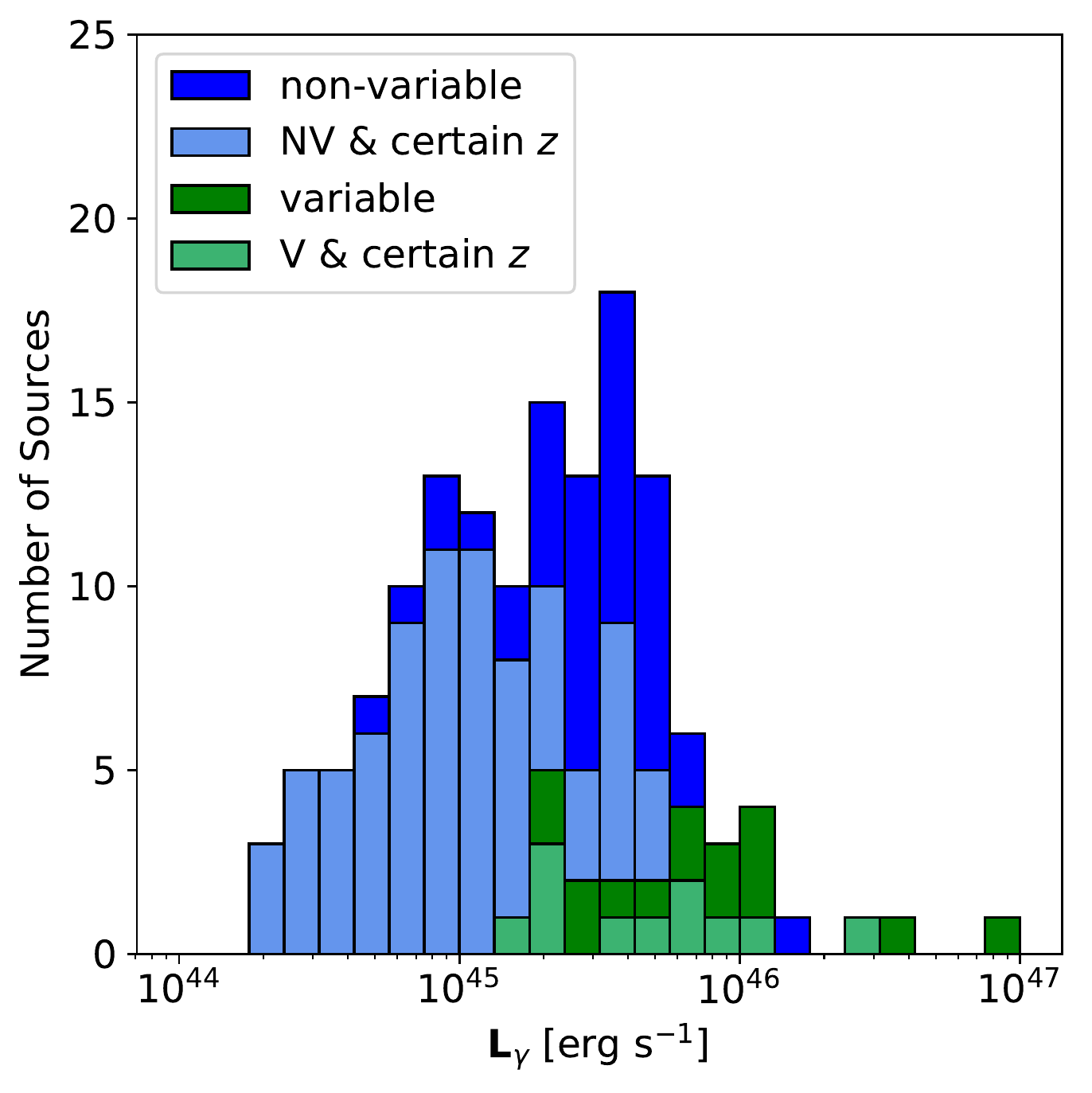}
\caption{Distribution of luminosity over the 1 - 300 GeV band for non--variable (blue) and variable (green) sources. Light color corresponds to sources with confident redshift measurements. Variable and non-variable sources are represented by ``V'' and ``NV'', respectively.}
\label{fig:lum_dist}
\end{center}
\end{figure}

The photon flux and luminosity distributions for non--variable and
variable sources are plotted in Figures \ref{fig:flux_dist} and
\ref{fig:lum_dist} respectively. For calculation of the luminosity we
adopt the following cosmology, $\Omega_{\Lambda} = 0.7$, $\Omega_{m} =
0.3$, $\Omega_{k} = 0$, and H$_0 = 70$ km/s/Mpc.

These distributions clearly show that our non-variable sources are less bright than those which exhibit stronger
variability. It is, however, less clear whether there is a correlation
between variability and source luminosity. Our result could simply 
be due to the fact that variability is more easily seen for nearby sources. 

\begin{figure}
\begin{center}
\includegraphics[width=\linewidth]{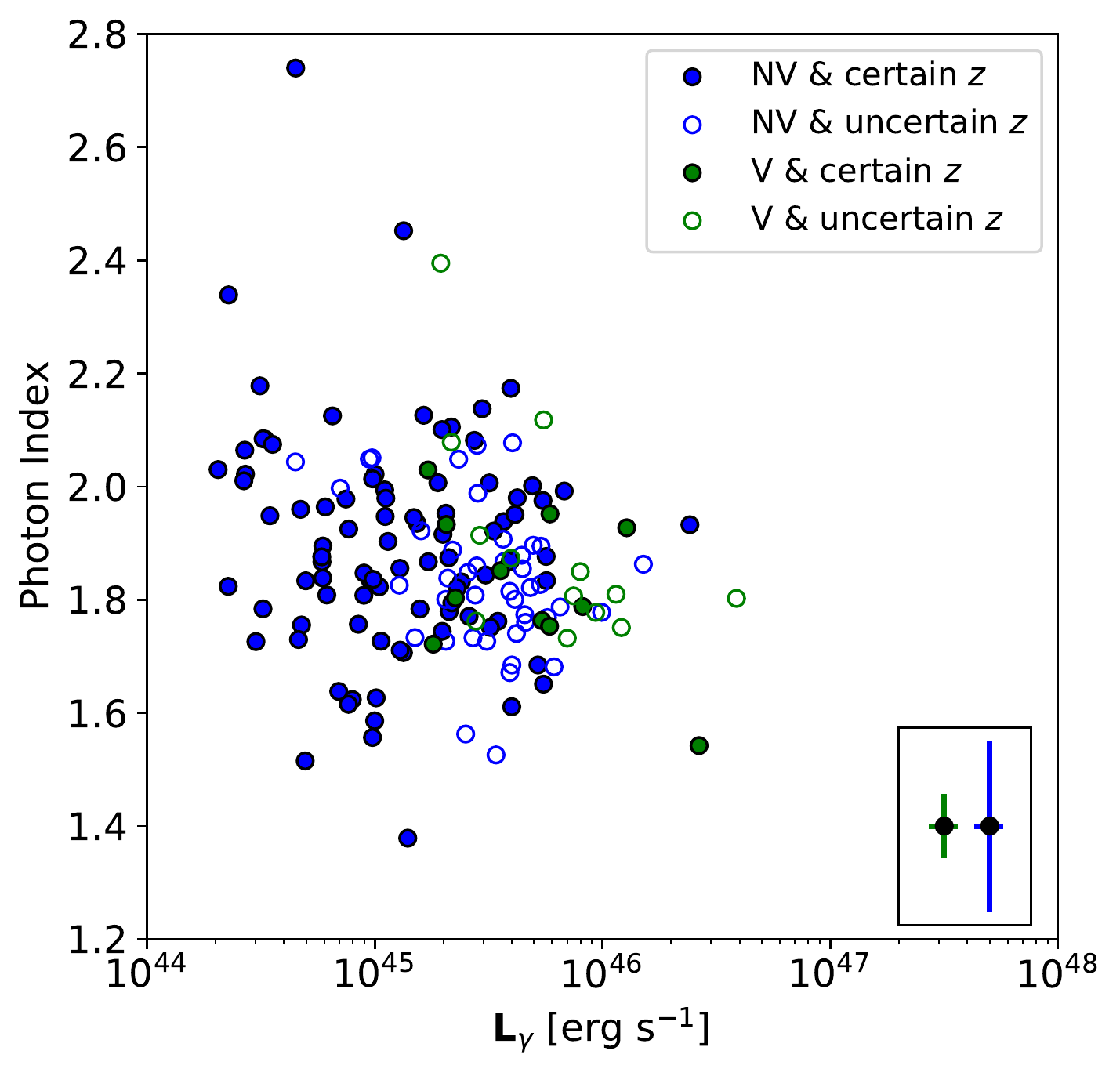}
\caption{Source luminosities against the photon index. Open circles
  correspond to sources of uncertain redshift. The legend at the lower
  right corner shows the characteristic error bars for sources of
  confident redshift. Note there is one source with photon index $< 1.20$ which is not shown here.}
\label{fig:photInd_v_lum}
\end{center}
\end{figure}

\begin{figure}
\begin{center}
\includegraphics[width=\linewidth]{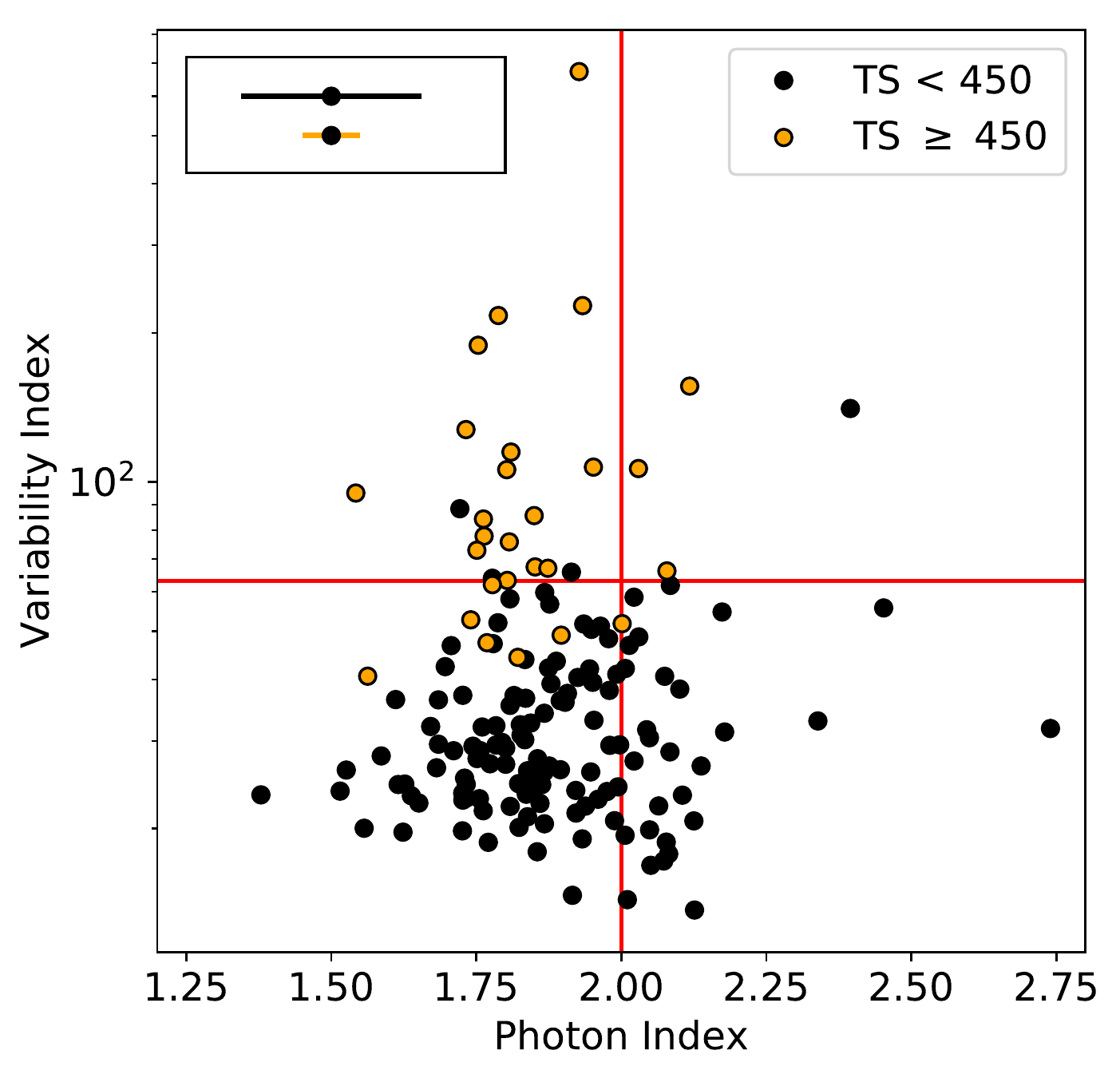}
\caption{Plot of the variability versus the photon index. Note that we have
  artificially split the data by choosing $TS\geq$ 450. This
  corresponds to an average $TS$ per temporal bin of 3. The red vertical line
    indicates a photon index $n=2$ and the red horizontal line indicates the 90\% confidence limit on source variability. Characteristic
  error bars for both classes are given in the legend of the upper
  left of the figure. Note there is one source with photon index $< 1.20$ which is not shown here.}
\label{fig:var_v_photInd}
\end{center}
\end{figure}

\begin{figure}
\begin{center}
\includegraphics[width=\linewidth]{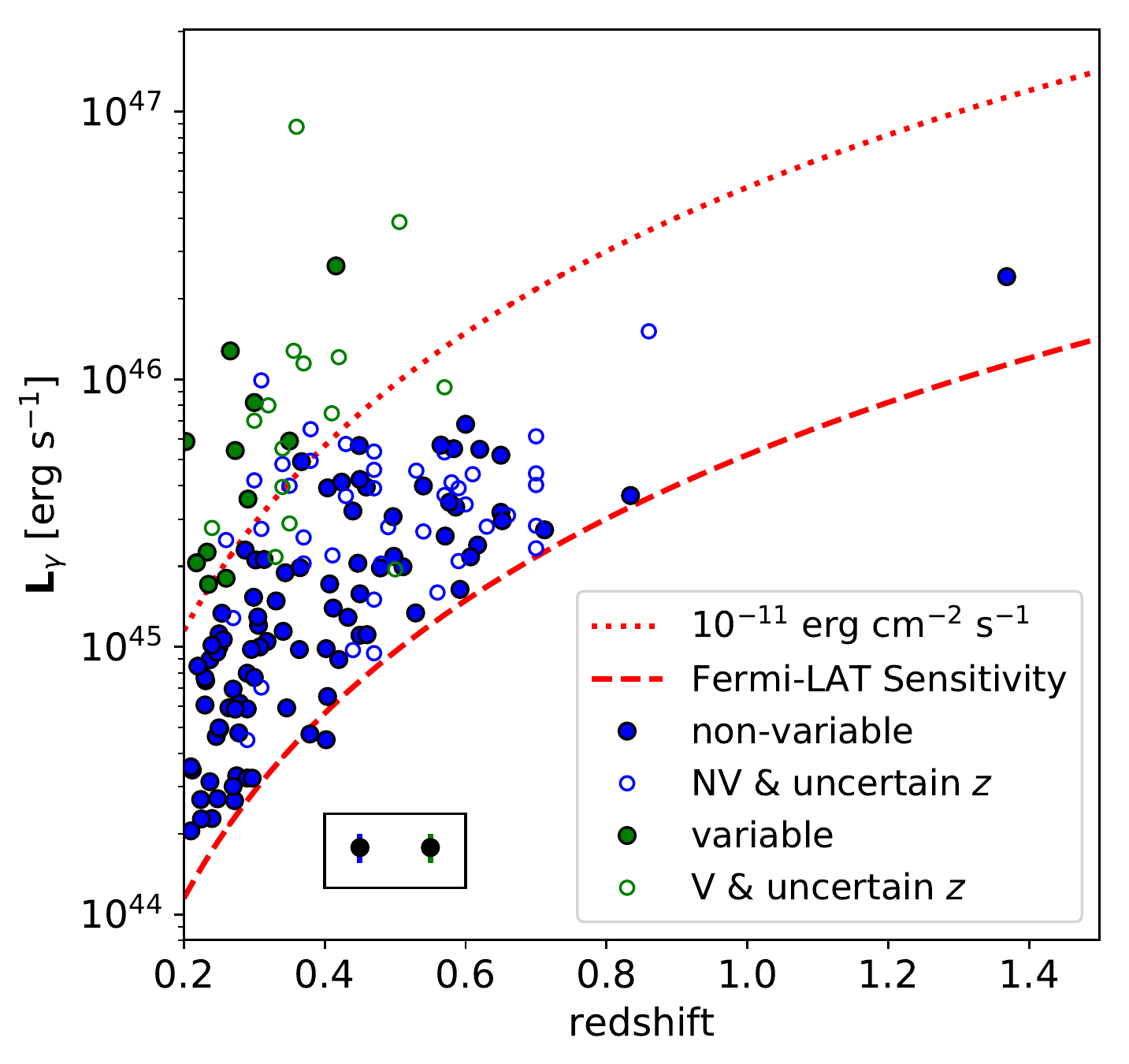}
\caption{Luminosity is plotted against redshift for the sources where
  open circles correspond to uncertain redshift. The thick dashed line
  corresponds to the 8 year, 5$\sigma$ {\it Fermi}-LAT detection
  threshold. Characteristic error bars are given in the legend on the
  lower left of the figure.}
\label{fig:lum_v_z}
\end{center}
\end{figure}

In Figure \ref{fig:photInd_v_lum} the measured photon index is plotted
against the calculated luminosity. For variable sources there appears
to be a weak trend between these two parameters. A correlation test on
the data reveals that the photon index is anti-correlated with source
luminosity at $\approx 2\sigma$ confidence. On the other hand, there
is no apparent correlation for non-variable data. Characteristic error
bars are depicted in Figure \ref{fig:photInd_v_lum} for variable and
non-variable sources which corresponds to the average error for each
class. Additionally plotted was variability index against the photon
index in Figure \ref{fig:var_v_photInd} to see if there was a
correlation. If one considers the data set as a whole, there is no
apparent correlation. Even further consideration of the strongest sources
implies there is no correlation between hardness and
variability. In Figure \ref{fig:var_v_photInd} we make this distinction
by considering sources with a $TS< 450$ as being sources with less
confident variability. Note that a non--variable source with $TS=
450$ with a variability index calculated over 50 time intervals will
have a test statistic for a per bin in the light-curve, corresponding to $\approx 3 \sigma$.

\begin{figure}
\begin{center}
\includegraphics[width=\linewidth]{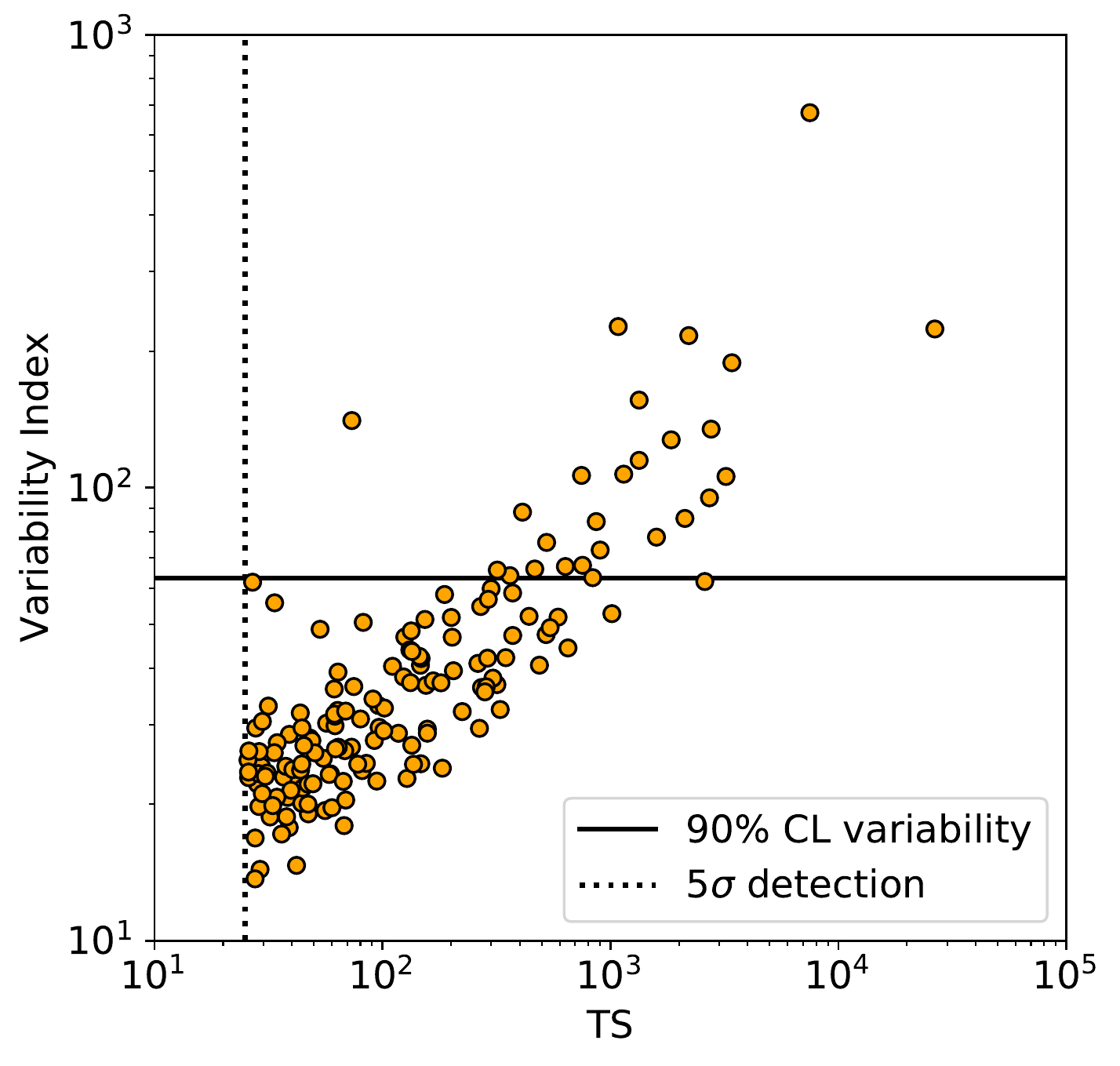}
\caption{Source variability index plotted against its $TS$ value. The solid black line indicates the 90\% confidence limit on source variability and the vertical dashed line the $TS \ge 25$, $\approx 5\sigma$, detection threshold. Note there are no sources below a $TS = 25$ as these did not meet the detection threshold.}
\label{fig:var_v_ts}
\end{center}
\end{figure}

Indeed, the true nature of the variability for each source should be
considered. It is more than likely that many of the sources from this
analysis do exhibit some variability, to which our analysis is not yet
sensitive. In Figure \ref{fig:lum_v_z}, where we plot the source
luminosity as a function of redshift, it is interesting to note that
the variable and non--variable sources can be roughly partitioned by
plotting the luminosity for a given energy flux over a range of
redshifts. In Figure \ref{fig:var_v_ts} we plot the variability as a
function of test statistic. There is, unsurprisingly, a strong
apparent correlation between the two.

\subsection{Promising Sources}

\begin{figure*}
  \label{fig_sources} 
  \begin{minipage}[b]{0.49\linewidth}
    \centering
    \includegraphics[width=\linewidth]{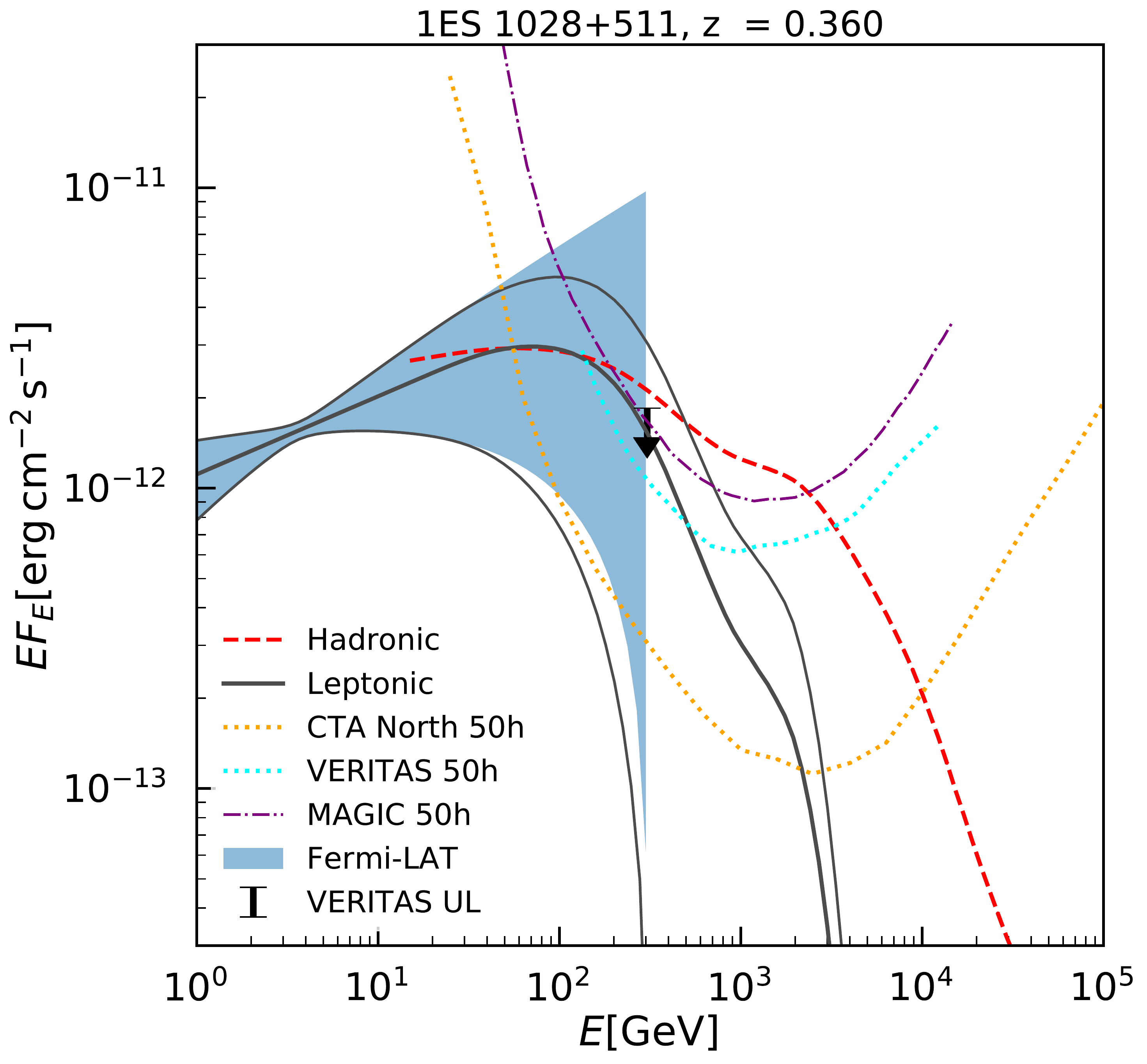} 
    \vspace{4ex}
  \end{minipage}%%
  \begin{minipage}[b]{0.49\linewidth}
    \centering
    \includegraphics[width=\linewidth]{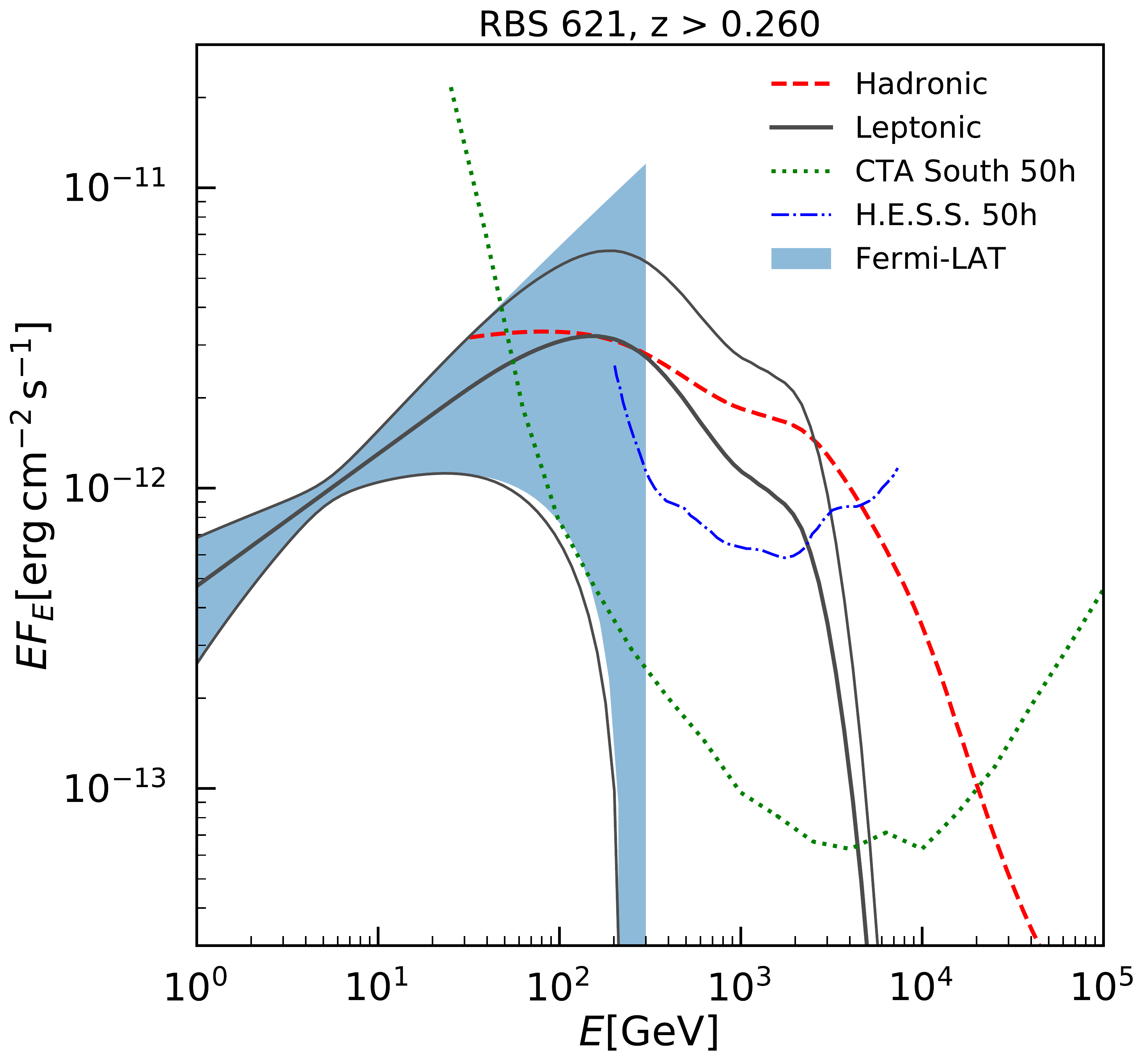}
    \vspace{4ex}
  \end{minipage} 
  \begin{minipage}[b]{0.49\linewidth}
    \centering
    \includegraphics[width=\linewidth]{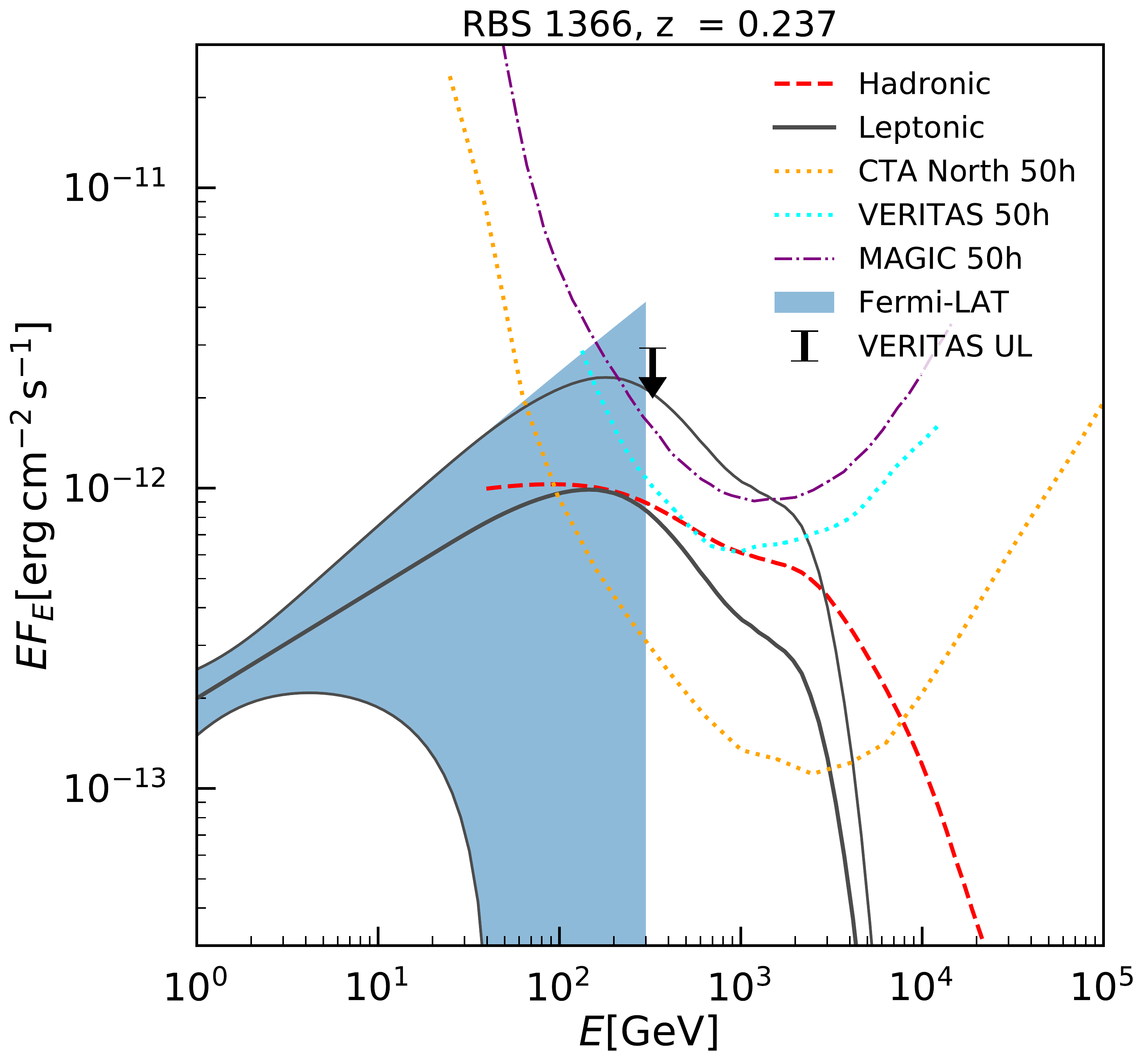} 
  \end{minipage}
    \begin{minipage}[b]{0.49\linewidth}
    \centering
    \includegraphics[width=\linewidth]{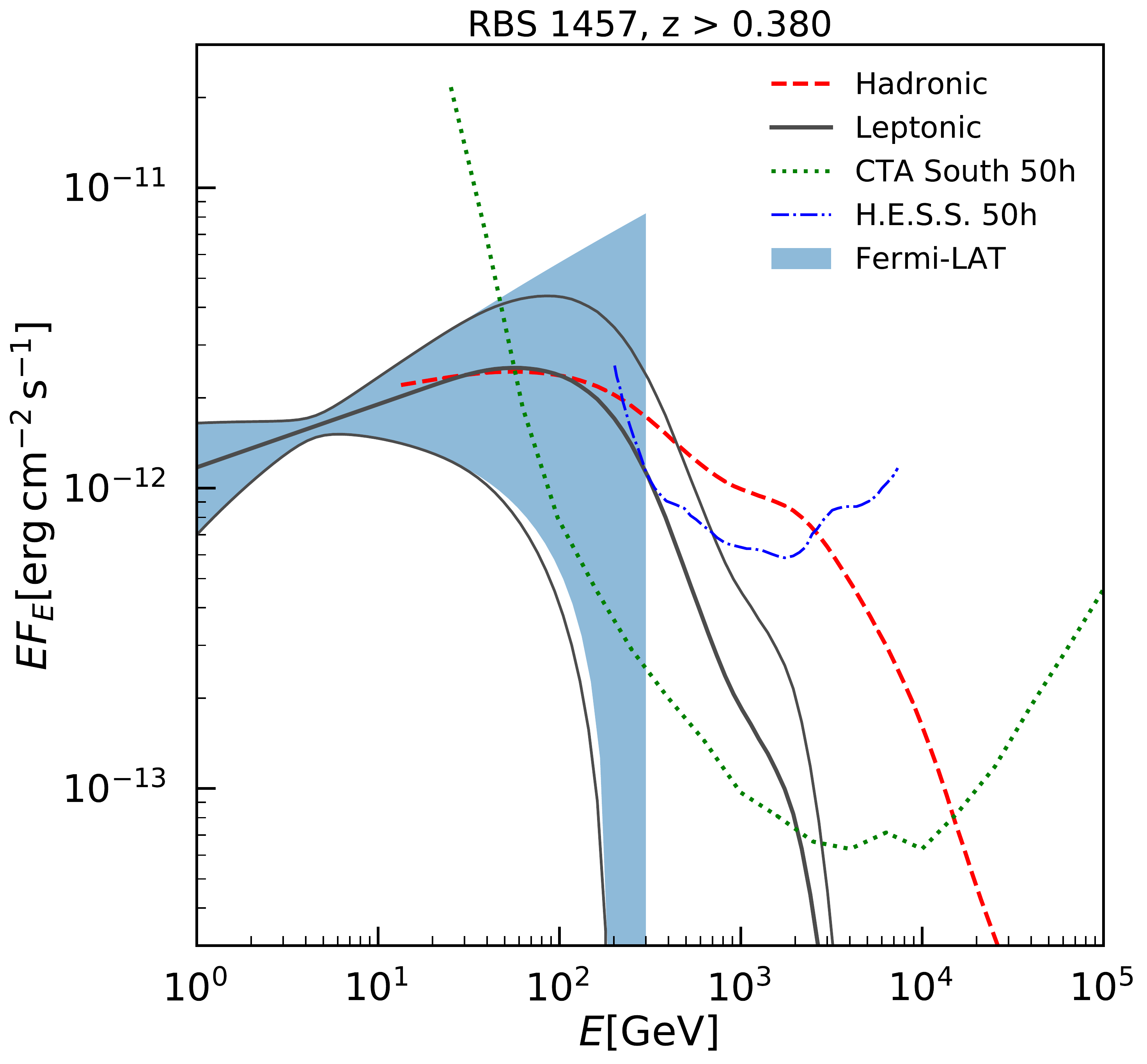} 
  \end{minipage}%% 
  
\caption{Modeled leptonic spectrum (thick black) with $1~\sigma$ confidence interval (grey) and the UHECR-induced cascade spectrum (solid red) of our promising sources based upon analysis of the textit{Fermi}-LAT data, shaded blue, in the GeV regime. Also plotted are the sensitivity curves for 50~h observations with detectors that can see the source. Sensitivities for \href{https://www.cta-observatory.org/science/cta-performance}{CTA} , North (yellow) and South (green), from \citep{2017APh....93...76H}, MAGIC (purple) \citep{2016APh....72...76A}, H.E.S.S. (blue), \citep{2015arXiv150902902H} and \href{http://veritas.sao.arizona.edu/about-veritas-mainmenu-81/veritas-specifications-mainmenu-111}{VERITAS} (cyan). Upper limits were calculated for the sources 1ES 1218+511 and RBS 1366 from observations with VERITAS \citep{2016AJ....151..142A}.}
\end{figure*}

From the results of the likelihood and variability analyses,
significant sources were discriminated based on their variability,
redshift, and brightness, to establish the best candidates with the
potential for a hadronic signature to be observed by IACTs, including
the Cherenkov Telescope Array (CTA)~\citep{2019scta.book.....C}, Major
Atmospheric Gamma Imaging Cherenkov Telescopes (MAGIC)
\citep{2016APh....72...61A}, High Energy Stereoscopic System
(H.E.S.S) \citep{2017A&A...600A..89H}, and Very Energetic Radiation Imaging Telescope Array System (VERITAS) \citep{2008AIPC.1085..657H}. 

We have identified four sources as Class I, based on the criterion that they may be detectable with current IACTs. We present them in turn, below.

\subsubsection{1ES 1028+511}

Source 2WHSP J103118.4+505335, 3FGL J1031.2+5053, also known with the name 1ES 1028+511, shown in 
Figure~\ref{fig_sources}, is a promising candidate for
observation by CTA and potentially MAGIC and VERITAS, based on our {\it Fermi}
analysis. This source has been observed by VERITAS for 24.1 hours, yielding an upper limit of of $7.7 \times 10^{-12} ~ {\rm cm^{-1} s^{-1} erg^{-1}}$~\citep{2016AJ....151..142A}. Interestingly, for this source the VERITAS upper limit constrains the hadronic model for the level of flux predicted with the best fit \Fermi index. Thus additional observations of this source may be very sensitive to or otherwise very constraining of the hadronic model. Additionally, with $\log \nu_{S} = 17.0$ it is interesting as a possible extreme-HSP source. 
 
\subsubsection{RBS 1366}

With a detection significance of $\approx 12\sigma$, a very low
variability index, and a firm redshift determination
\citep{2012ApJS..203...21A}, 2WHSP J141756.5+254324, 3FGL J1417.8+2540, also known as RBS 1366, is also a promising candidate for detecting with TeV instruments. The large uncertainties
in the {\it Fermi}-LAT analysis however mean that we cannot
conclusively determine whether the source will be detectable. Under
optimistic assumptions (upper $1\sigma$ uncertainty range) it is a
good candidate for observation by CTA or VERITAS and also possibly by MAGIC. 
 A differential upper limit of $1.7 \times 10^{-11} ~ {\rm cm^{-1} s^{-1} erg^{-1}}$ at 327 GeV was calculated for this source by VERITAS based on 10 hours of observations, but is not sufficient to constrain the hadronic origin model~\citep{2016AJ....151..142A}.
With synchrotron peak frequency at $\log \nu_{S} = 17.4$, 
this source is possibly an extreme-HSP, and thus interesting to study at VHE even if it is purely leptonic, for the purpose of furthering our knowledge of this, 
small and extreme source population. It was also flagged as a TeV blazar candidate by the analysis of~\citet{Costamante_2019}.  

\subsubsection{RBS 621}

The blazar RBS 621 (3FGL J0506.9-5435, 2WHSP J050657.7-543503), with \Fermi-LAT detection significance of around 23.0 $\sigma$ and low variability TS$_V = 40.6$, is another source promising for TeV detection and for the detection of the UHECR hadronic signature. The redshift is uncertain with a lower limit of $ z > 0.26$. To the best of our knowledge this source has not yet been observed with H.E.S.S., but it is our most promising source for detection with a 50h exposure if the true redshift is close to the lower limit and certainly promising for observations with CTA. 

\subsubsection{RBS 1457}

The blazar RBS 1457 (3FGL J1503.7-1540,2WHSP J150340.6-154113) is also one of the sources most promising for TeV detection and for detection
of the hadronic cascade signature in our sample, with \Fermi-LAT detection significance of around 21.0 $\sigma$ and low variability TS$_V = 52$. There is only a lower limit on the redshift of this source, $ z > 0.38$, but if the true redshift is not much higher than the lower-limit, this source, at declination $\delta = - 15.4^{\circ}$ is possibly detectable with H.E.S.S. and it is certainly a promising source for CTA South. 

\subsubsection{Other promising sources}

We have found a number of additional promising sources, which we categorised as Class II because they likely require an instrument with sensitivity comparable to that of CTA for detection. We show their \gRay spectra in Figure~\ref{fig_sources_ii}, in the Appendix. The source 2WHSP J143657.7+563924, or RBS 1409, or RX J1436.9+5639, is one of these sources. An upper limit was obtained based on a 13h observation of the source with VERITAS~\citep{Aliu_2012}. A redshift of $z = 0.15$ has been quoted based on the redshift of a galaxy cluster within the same region of the sky~\citep{Bauer_2000}. However, the optical spectrum of the galaxy is featureless~\citep{Aliu_2012}. We thus assumed a redshift value  equal to the lower limit quoted by the more recent work of ~\citet{Chang:2016mqv} in the present work. The Class II sample additionally contains Ton 116, which at redshift $z = 1.065$, if detected with CTA could give unambigious evidence of the hadronic cascade. The additional notable candidates include 3FGL J1124.9+4932 - RBS 981, at redshift $z >0.57$, and RGB 1756+553 at redshift $z >0.57$, both suitable with observations with North sky instruments, as well as 3FGL J0118.9–145 at redshift $z > 0.530$ and 1ES 0602-482 in the Southern sky. 

Two sources from Table \ref{tab:example_table}, 3FGL J2055.2-0019 and 3FGL J1012.7+4229, are not included as promising sources for observation even though they met the initial selection criteria. It was clear that current and future IACTs lack the sensitivity to detect the hadronic componant of these sources in a reasonable observation period.

\section{Discussion and Conclusions}
\label{sec:discussion}
We have analyzed the {\it Fermi} data for 566 HSP blazars from the 2WHSP
catalog. By discriminating significant sources with {\it Fermi}-LAT
data in the GeV band based on the hardness of spectrum, limited
variability, and detectability with current and future IACTs, we have
compiled a list of the most promising sources for TeV follow-up
observations. By extending the GeV spectrum to greater than TeV
energies and modelling the expected \gray{} spectrum 
under the assumption of both leptonic and hadronic origins, we
have shown that if the sources are UHECR accelerators, and 
magnetic fields in the intergalactic medium are low,
it will be possible to distinguish between a leptonic or UHECR-induced
cascade scenario in these sources using CTA.

The motivation for our analysis has been twofold. Firstly, to identify
blazars whose \gRay spectra can be used, if a detection with IACTs is
achieved, to constrain the extragalactic magnetic field by considering
their combined GeV-TeV spectra~\citep[e.g.,][]{Murase08b,NS09,2010Sci...328...73N,Tavecchio:2010mk,Dermer11,Dolag11,Taylor11}. Secondly,
to identify blazars which could exhibit a hard spectrum in the TeV
energy range, which could be the signature of UHECR acceleration and
emission from these sources, as has been previously discussed for a
handful of extreme
HSPs~\citep[e.g.,][]{2010APh....33...81E,Essey:2010er,Murase:2011cy,Takami:2013gfa,Aharonian13,Tavecchio14,Oikonomou:2014xea,Dzhatdoev:2016ftt,2019MNRAS.483.1802T,2019scta.book.....C,Khalikov:2019fbd}. Recently
the MAGIC Collaboration announced the detection of 2WHSP
J073326.7+515354~\citep{Acciari:2019ntl} at $\gtrsim1$~TeV energies. The
source does not form part of our sample, since it lies at $z = 0.06$
and doesn't satisfy the $z \geq 0.2$ cut. However, the search for TeV
emission from this source demonstrates the interest for detectable HSP
and extreme HSP sources, as well as some of the open questions on
blazar emission which can be addressed with similar IACT observations.

In our analysis the majority of sources were found to lack any
significant variability. While blazars as a class are well known for
their variability, this result cannot exclude the presence of
variability, since these are relatively faint sources, and any
intrinsic variability may be below the experimental sensitivity of
{\it Fermi}. As expected in this statistically limited regime, the
sources that do exhibit significant variability in our analysis are
the brightest sources in terms of flux. Therefore, our results are not
conclusive in regards to the question whether some of the sources
examined could be powered by secondary \gRays from UHECR primaries. In
the latter case, there should be no detectable variability in the
\gRay energy range since UHECRs get delayed by magnetic fields. Given
that the bright sources in our sample are all consistent with being
variable, our results are consistent with all sources being
intrinsically variable and thus powered by leptonic emission
mechanisms, giving no conclusive support to the UHECR-induced cascade scenario.

It is important to emphasise again how our variability results differ
from those which accompany the 4FGL catalog. The most important
difference is the energy range over which the variability index was
calculated. In our analysis we used a higher
energy threshold of 1 GeV than the 100 MeV threshold of the 4FGL, which is better suited for the hard spectrum sources of interest in our analysis and might allow us, to better isolate the the hadronic component which, if present, should be dominant above $\sim10-100$ GeV as demonstrated in previous sections. The present analysis was conducted
using data from the {\it Fermi} launch to late 2016. An updated
analysis should reach the same conclusion, but surely with improved
uncertainties ( $\approx$ 10\% with 2 more years of data).

While the present work was being finalised, the 3HSP catalogue, which
is the largest and most complete HSP catalogue available to date,
became available~\citep{2019arXiv190908279C}. With respect to the
2WHSP it contains 395 additional HSP blazars. In the future, our analysis
could be extended to include these additional sources.

\section*{Acknowledgements}
This work was initiated in 2016 for the honors thesis ``BLAZARS AS A SOURCE FOR ULTRA-HIGH-ENERGY COSMIC RAYS? A SEARCH FOR THE ELUSIVE HADRONIC SIGNATURE'' at the Schreyer Honors College of Penn State University in 2018. 
We thank Gordana Te\v{s}i\'{c} for fruitful input during the early stages of this work.
We also thank Akira Okumura for useful comments on sensitivity curves. 
The work of K.M. is supported by the Alfred P. Sloan Foundation, NSF Grant No.~AST-1908689, and KAKENHI No.~20H01901.
%%%%%%%%%%%%%%%%%%%%%%%%%%%%%%%%%%%%%%%%%%%%%%%%%%



\bibliographystyle{mnras}
\bibliography{TOM.bib}

%%%%%%%%%%%%%%%%%%%%%%%%%%%%%%%%%%%%%%%%%%%%%%%%%%

%%%%%%%%%%%%%%%%% APPENDICES %%%%%%%%%%%%%%%%%%%%%

\appendix


%%%%%%%%%%%%%%%%%%%%%%%%%%%%%%%%%%%%%%%%%%%%%%%%%%
\newpage
\onecolumn

%\section*{Additional Figures}

\begin{figure*}
  \centering
  \label{fig_sources_ii} 
  \begin{minipage}[b]{0.35\linewidth}
    \centering
    \includegraphics[width=\linewidth]{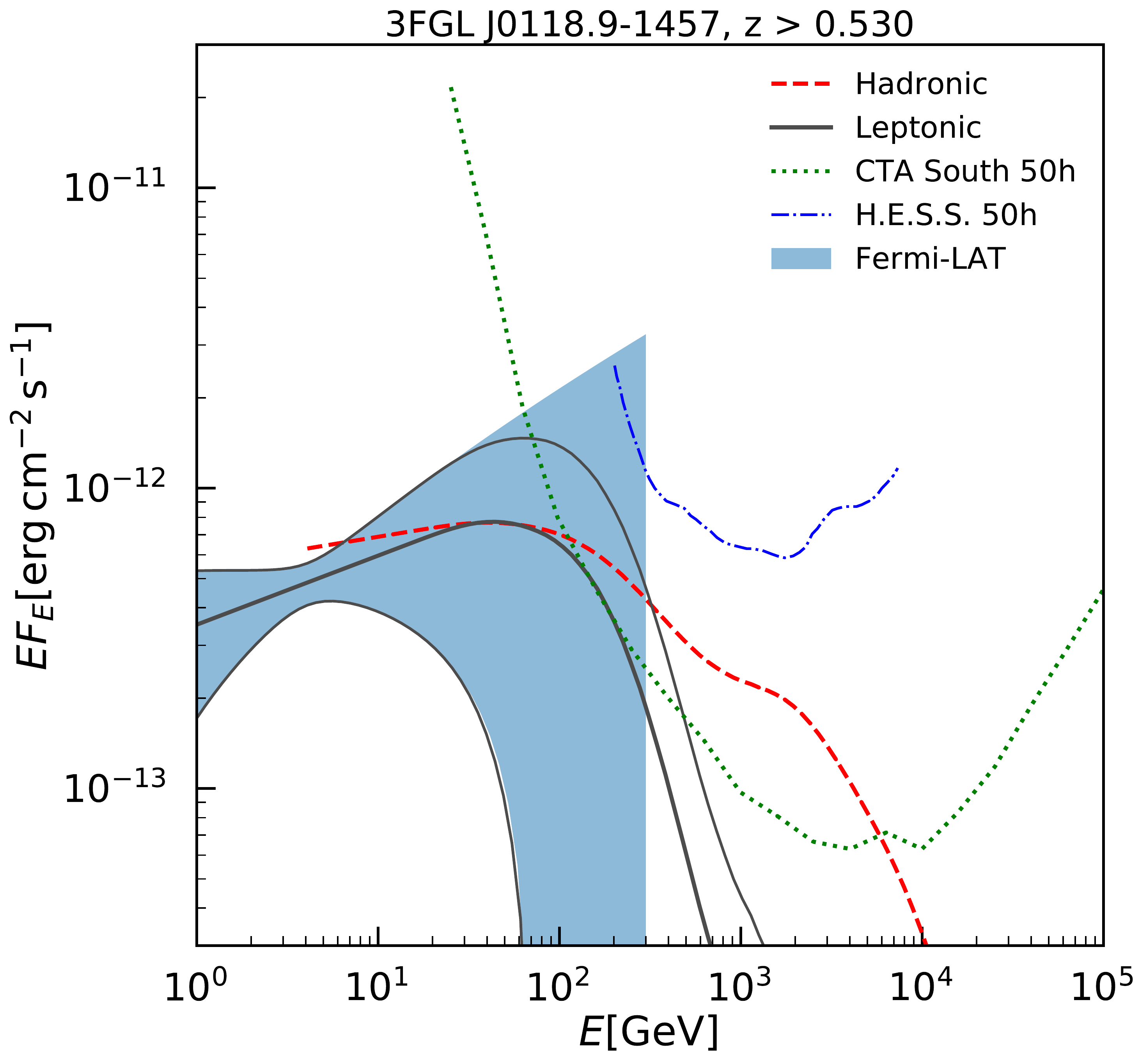} 
    \vspace{4ex}
  \end{minipage}%%
  \begin{minipage}[b]{0.35\linewidth}
    \centering
    \includegraphics[width=\linewidth]{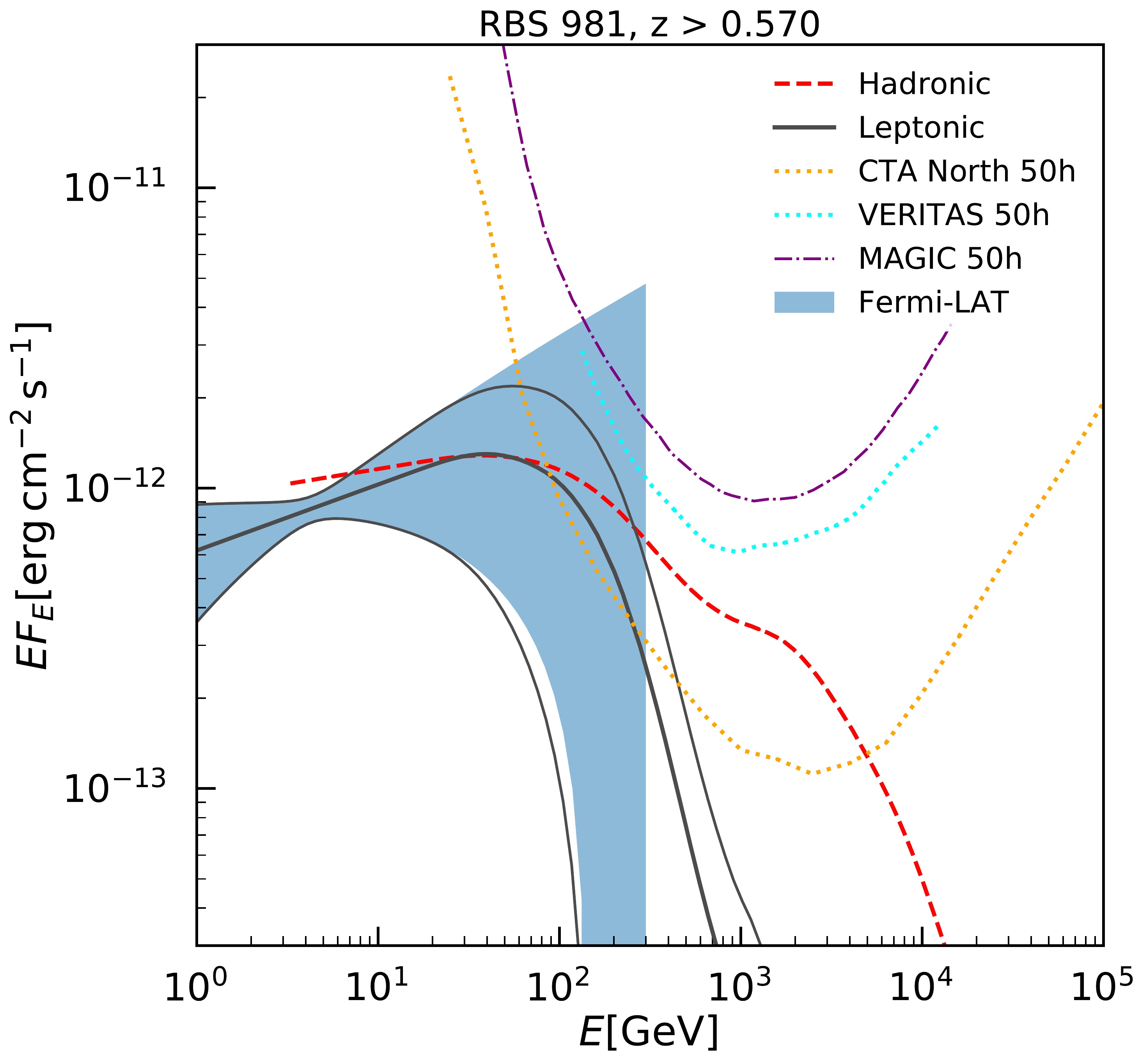} 
    \vspace{4ex}
  \end{minipage} 
  \begin{minipage}[b]{0.35\linewidth}
    \centering
    \includegraphics[width=\linewidth]{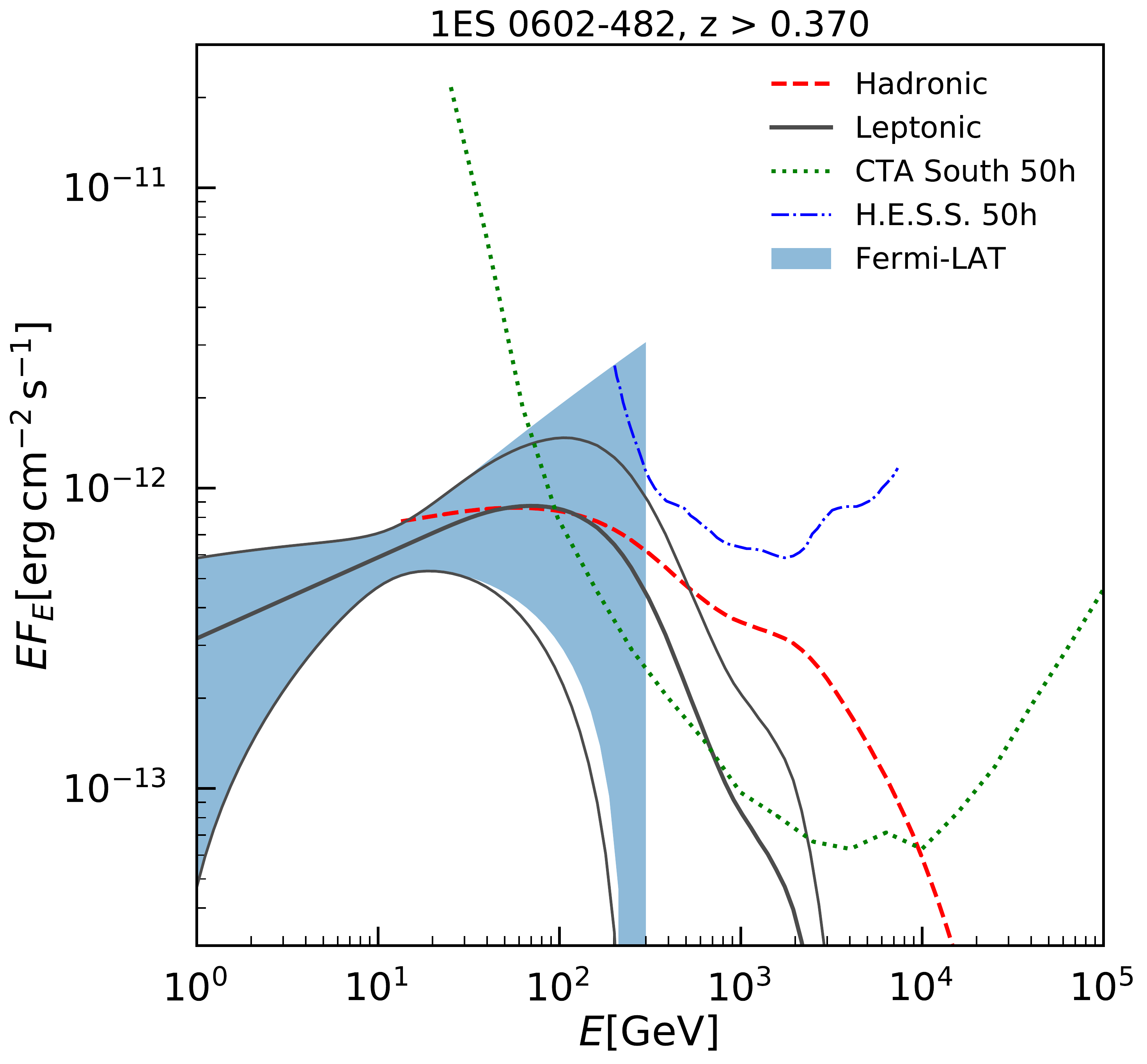} 
  \end{minipage}%% 
  \begin{minipage}[b]{0.35\linewidth}
    \centering
    \includegraphics[width=\linewidth]{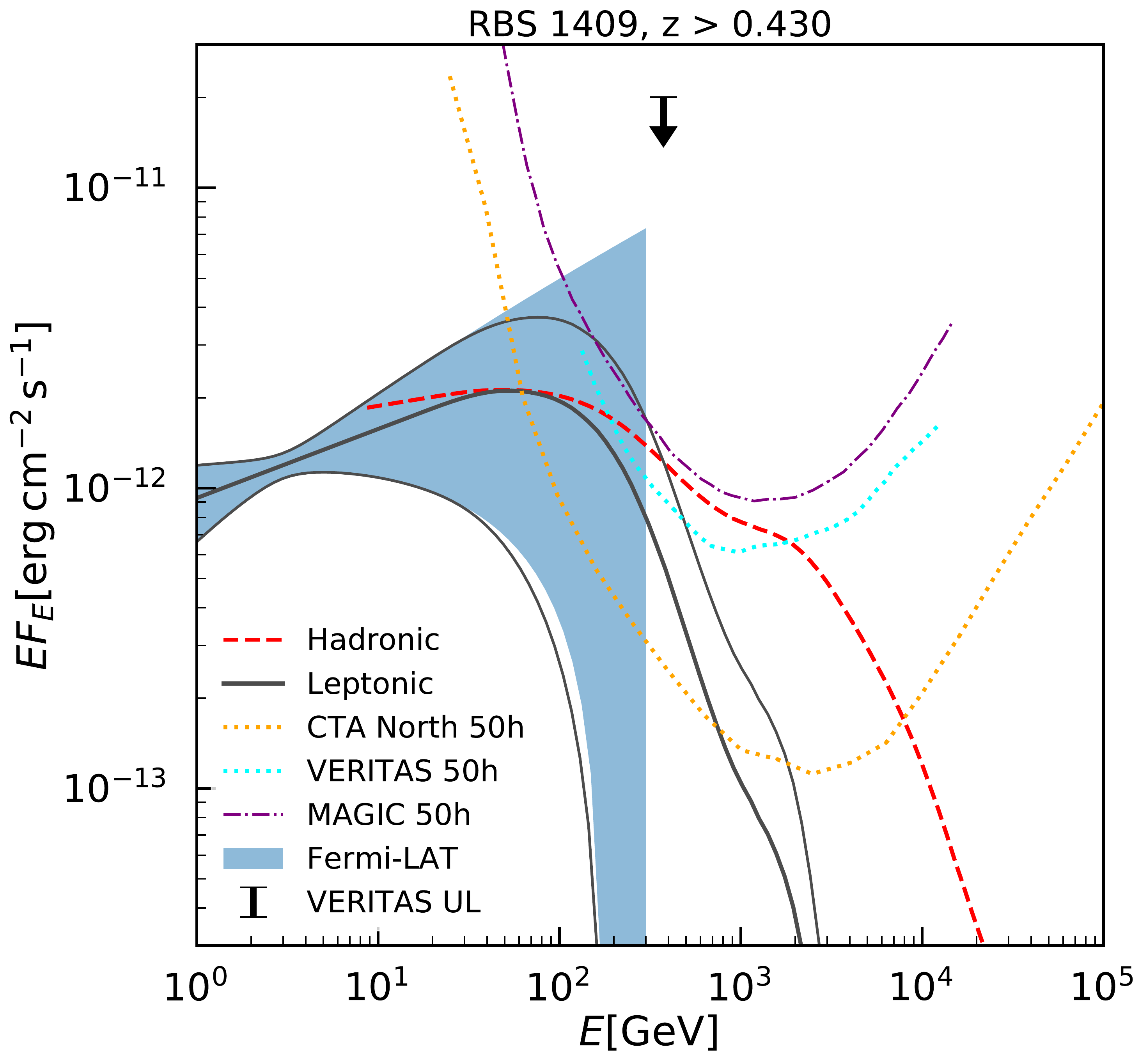} 
  \end{minipage}
  \begin{minipage}[b]{0.35\linewidth}
    \centering
    \includegraphics[width=\linewidth]{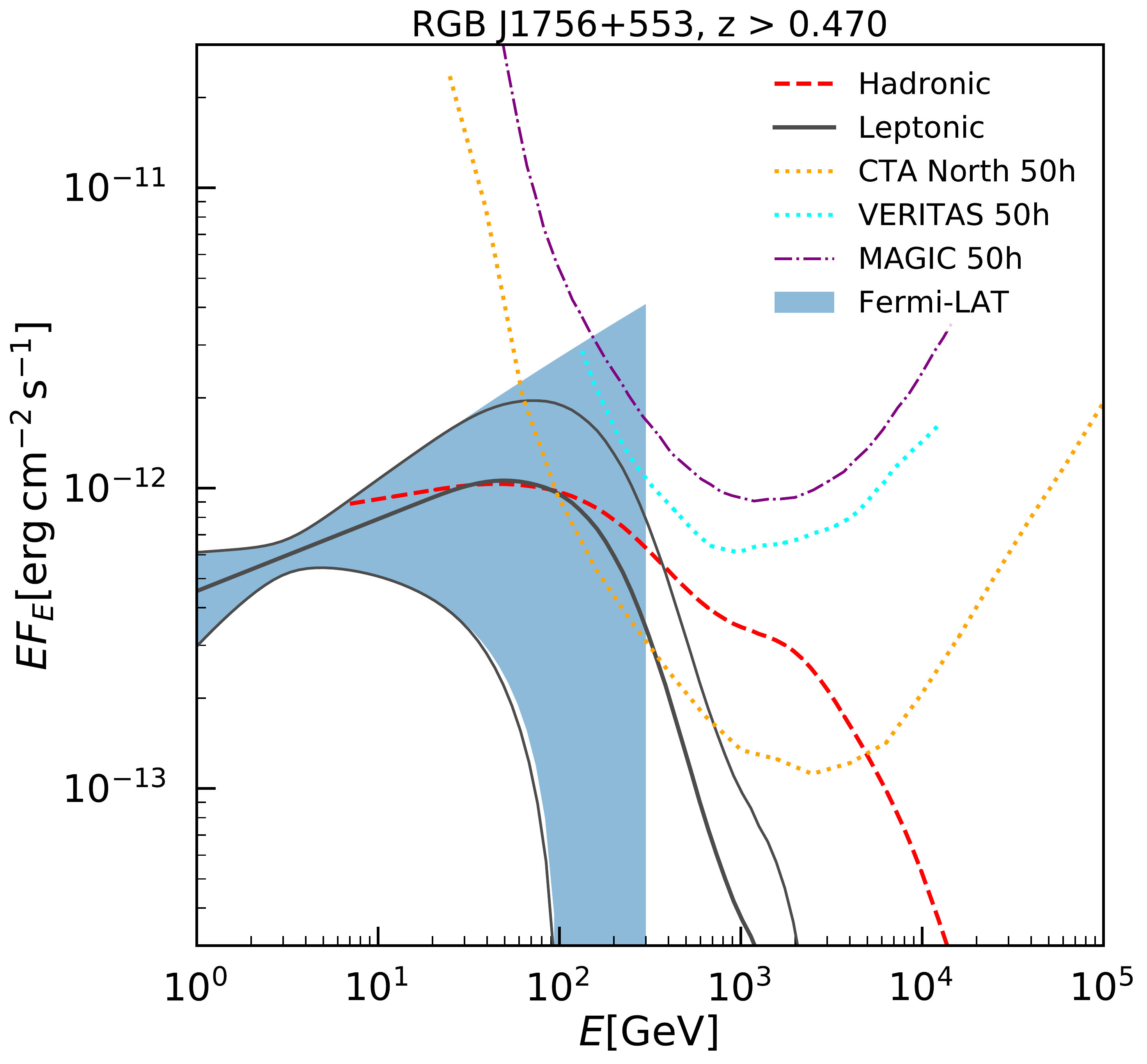} 
  \end{minipage}
  \begin{minipage}[b]{0.35\linewidth}
    \centering
    \includegraphics[width=\linewidth]{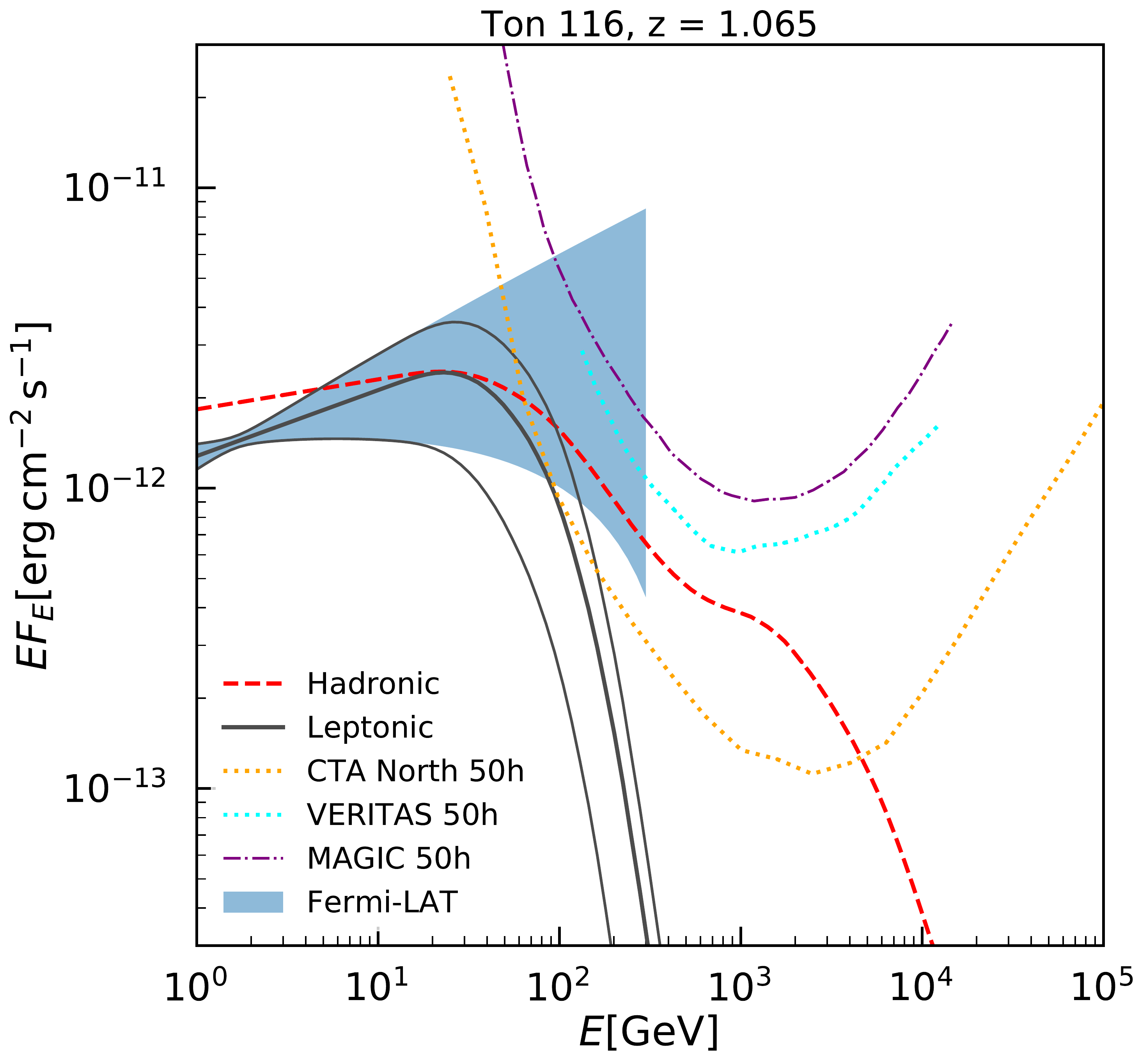} 
  \end{minipage}%% 
\caption{Modeled leptonic spectrum (thick black) with $1~\sigma$ confidence interval (grey) and the UHECR-induced cascade spectrum (solid red)  of our promising sources based upon analysis of the textit{Fermi}-LAT data, shaded blue, in the GeV regime. Also plotted are the sensitivity curves for 50~h observations with detectors that can see the source. Sensitivities for \href{https://www.cta-observatory.org/science/cta-performance}{CTA} , North (yellow) and South (green), from \citep{2017APh....93...76H}, MAGIC (purple) \citep{2016APh....72...76A}, H.E.S.S. (blue), \citep{2015arXiv150902902H} and \href{http://veritas.sao.arizona.edu/about-veritas-mainmenu-81/veritas-specifications-mainmenu-111}{VERITAS} (cyan). Upper limits were calculated for the source RBS 1409 from observations with VERITAS \citep{Aliu_2012}.}
\end{figure*}

\pagebreak
\clearpage

\newpage
\section*{Table 1 (Complete)} 
\scriptsize
\LTcapwidth=1.0\textwidth
\begin{longtable}{ccccccccccccc}
\caption*{\normalsize Results from the analysis of 2WHSP sources. Included is the name from the 2WHSP catalog, the luminosity from 1 -- 300 GeV,  $\boldmath L_{44}$, in units of $10^{44}$ erg s$^{-1}$, the 1 -- 300 GeV photon flux, $\boldmath {\left({\rm d}N/{\rm d}t\right)}_{-10}$, in units of $10^{-10}$ cm$^{-2}$ s$^{-1}$, the test statistic from the likelihood fit , $\boldmath{TS}$, the normalization and its error scaled by $10^{-11}$ cm$^{-2}$ s$^{-1}$ GeV$^{-1}$, $N_{-11}$ and $\sigma_{N}$, the photon index $\gamma$ and its error $\sigma_{\gamma}$, the variability index $TS_V$ (> 63.17 is a variable source), the source redshift $z$, a statement on the uncertainty of the redshift ($T$ means redshift is uncertain), and alternative identifier, and the logarithm of the synchrotron peak frequency, $\log \nu_{\rm S, pk}$ in units of $\log{\text{Hz}}$.} \\

\hline \textbf{2WHSP Name} & \boldmath$L_{44}$ & \boldmath${\left(\frac{dN}{dt}\right)}_{-10}$ & \boldmath$TS$ & \boldmath$N_{-11}$ & \boldmath$\sigma_{N}$ & \boldmath$\gamma$ & \boldmath$\sigma_{\gamma}$ & \boldmath$TS_V$ & \textbf{z} & \textbf{$U_z$} &\textbf{Other Name} & \textbf{$\log \nu_{\rm S, pk}$} \\ \hline
\endfirsthead

\multicolumn{13}{c}%
{\tablename\ \thetable{} -- continued from previous page} \\
\hline \textbf{2WHSP Name} & \boldmath$L_{44}$ & \boldmath${\left(\frac{dN}{dt}\right)}_{-10}$ & \boldmath$TS$ & \boldmath$N_{-11}$ & \boldmath$\sigma_{N}$ & \boldmath$\gamma$ & \boldmath$\sigma_{\gamma}$ & \boldmath$TS_V$ & \textbf{z} & \textbf{$U_z$} &\textbf{Other Name} & \textbf{$\log \nu_{\rm S, pk}$}\\ \hline
\endhead

\hline \multicolumn{13}{|r|}{{Continued on next page}} \\ \hline
\endfoot

\hline
\endlastfoot
J002200.9+000657    &    12.03    &    0.66    &    40.37    &    0.5    &    0.37    &    1.16    &    0.27    &    22.84    &                     0.306    &                    &   --   &         16.3  \\
J020412.9-333339    &    24.01    &    1.17    &    29.49    &    1.31    &    0.41    &    1.83    &    0.21    &    24.26    &                     0.617    &                    &   1BIGB J020412.9-3333   &         17.9  \\
J023536.6-293843    &    31.01    &    1.02    &    25.84    &    1.12    &    0.44    &    1.73    &    0.24    &    23.57    &    >                0.660    &                    &   --   &         15.8  \\
J030103.7+344100    &    2.28    &    2.43    &    31.57    &    2.5    &    0.57    &    2.34    &    0.23    &    32.96    &                     0.240    &                    &   1BIGB J030103.7+3441   &         15.7  \\
J030433.9-005403    &    19.92    &    1.82    &    41.97    &    2.05    &    0.58    &    1.92    &    0.2    &    14.66    &                     0.511    &                    &   1BIGB J030433.9-0054   &         15.4  \\
J031423.8+061955    &    54.73    &    3.53    &    81.43    &    3.94    &    0.75    &    1.98    &    0.15    &    23.74    &                     0.620    &    T               &   1BIGB J031423.8+0619   &         16.3  \\
J035856.1-305447    &    31.82    &    1.94    &    55.99    &    2.16    &    0.46    &    2.01    &    0.18    &    19.38    &                     0.650    &    T               &   1BIGB J035856.1-3054   &         16.9  \\
J050335.3-111506    &    36.91    &    2.35    &    68.98    &    2.64    &    0.62    &    1.87    &    0.16    &    20.44    &    >                0.570    &                    &   1BIGB J050335.3-1115   &         16.9  \\
J060714.2-251859    &    3.3    &    1.77    &    39.03    &    1.95    &    0.46    &    2.08    &    0.2    &    28.54    &                     0.275    &                    &   1BIGB J060714.2-2518   &         17.5  \\
J062149.6-341148    &    13.39    &    2.56    &    33.64    &    2.51    &    0.57    &    2.45    &    0.25    &    55.71    &                     0.529    &                    &   1BIGB J062149.6-3411   &         17.7  \\
J062337.7-525756    &    9.73    &    1.65    &    27.64    &    1.82    &    0.51    &    2.05    &    0.25    &    16.85    &    >                0.440    &                    &   --   &         15.2  \\
J062626.2-171045    &    40.25    &    2.33    &    37.95    &    2.56    &    0.65    &    2.08    &    0.22    &    18.77    &    >                0.700    &                    &   1BIGB J062626.2-1710   &         16.6  \\
J073152.6+280432    &    2.71    &    1.65    &    34.55    &    1.83    &    0.47    &    2.02    &    0.22    &    27.36    &                     0.248    &                    &   1BIGB J073152.6+2804   &         17.0  \\
J074929.5+745143    &    21.69    &    1.86    &    59.17    &    2.04    &    0.4    &    2.11    &    0.18    &    23.35    &                     0.607    &    T               &   --   &         16.1  \\
J083724.5+145819    &    4.77    &    1.29    &    36.82    &    1.44    &    0.46    &    1.76    &    0.2    &    22.97    &                     0.278    &                    &   1BIGB J083724.5+1458   &         16.7  \\
J090953.2+310602    &    2.66    &    1.29    &    29.04    &    1.43    &    0.43    &    2.01    &    0.25    &    14.37    &                     0.272    &                    &   1BIGB J090953.2+3106   &         17.0  \\
J095214.6+393615    &    23.35    &    1.28    &    29.72    &    1.42    &    0.41    &    2.05    &    0.25    &    30.48    &    >                0.700    &                    &   1FGL J0952.2+3926   &         16.5  \\
J095507.9+355100    &    36.78    &    1.07    &    28.26    &    1.2    &    0.39    &    1.94    &    0.25    &    22.19    &                     0.834    &                    &   1BIGB J095507.9+3551   &         17.5  \\
J100612.1+644010    &    15.96    &    1.19    &    39.66    &    1.33    &    0.36    &    1.92    &    0.19    &    21.49    &    >                0.560    &                    &   --   &         15.4  \\
J101706.6+520247    &    4.72    &    0.96    &    25.88    &    1.07    &    0.33    &    1.96    &    0.24    &    22.91    &                     0.379    &                    &   --   &         15.8  \\
J103346.3+370824    &    20.53    &    2.79    &    95.99    &    3.12    &    0.54    &    1.95    &    0.14    &    33.06    &                     0.447    &                    &   --   &         17.1  \\
J104857.6+500945    &    4.5    &    2.12    &    43.59    &    1.82    &    0.44    &    2.74    &    0.32    &    31.82    &                     0.402    &                    &   1BIGB J104857.6+5009   &         17.4  \\
J110357.1+261117    &    27.38    &    1.53    &    38.98    &    1.68    &    0.42    &    2.08    &    0.22    &    17.77    &                     0.712    &    T               &   --   &         17.9  \\
J113444.6-172900    &    25.94    &    1.33    &    32.13    &    1.49    &    0.55    &    1.77    &    0.23    &    18.76    &                     0.571    &                    &   1BIGB J113444.6-1729   &         16.9  \\
J121158.6+224233    &    11.05    &    1.6    &    37.65    &    1.78    &    0.46    &    1.99    &    0.21    &    24.27    &                     0.450    &                    &   1BIGB J121158.6+2242   &         17.6  \\
J122307.2+110038    &    241.85    &    2.04    &    47.3    &    2.29    &    0.52    &    1.93    &    0.17    &    19.05    &                     1.368    &    T               &   --   &         16.1  \\
J124141.4+344029    &    28.34    &    1.39    &    34.42    &    1.56    &    0.42    &    1.99    &    0.22    &    20.74    &    >                0.700    &                    &   1BIGB J124141.4+3440   &         16.6  \\
J125847.9-044744    &    33.29    &    2.22    &    40.44    &    2.48    &    0.69    &    1.92    &    0.2    &    23.87    &                     0.586    &    T               &   1BIGB J125847.9-0447   &         16.9  \\
J130145.6+405623    &    29.61    &    2.25    &    73.12    &    2.45    &    0.44    &    2.14    &    0.17    &    26.74    &                     0.652    &                    &   1BIGB J130145.6+4056   &         15.9  \\
J131234.6-185900    &    28.13    &    2.09    &    36.1    &    2.3    &    0.58    &    2.07    &    0.23    &    17.2    &    >                0.630    &                    &   --   &         16.0  \\
J133102.8+565541    &    3.01    &    0.82    &    28.69    &    0.9    &    0.33    &    1.73    &    0.21    &    19.78    &                     0.270    &                    &   --   &         17.6  \\
J135328.0+560056    &    6.52    &    1.53    &    38.42    &    1.67    &    0.4    &    2.13    &    0.22    &    20.72    &                     0.404    &                    &   1BIGB J135328.0+5600   &         16.3  \\
J142421.1+370552    &    3.23    &    1.54    &    26.92    &    1.69    &    0.54    &    2.08    &    0.29    &    61.85    &                     0.290    &                    &   --   &         16.3  \\
J152913.5+381216    &    20.92    &    1.15    &    29.7    &    1.29    &    0.38    &    1.84    &    0.2    &    21.12    &    >                0.590    &                    &   1BIGB J152913.5+3812   &         15.7  \\
J160218.0+305108    &    9.47    &    1.36    &    33.01    &    1.51    &    0.4    &    2.05    &    0.22    &    19.87    &    >                0.470    &                    &   1BIGB J160218.0+3051   &         15.6  \\
J164220.2+221143    &    16.4    &    1.55    &    27.61    &    1.69    &    0.51    &    2.13    &    0.26    &    13.69    &                     0.592    &                    &   1BIGB J164220.2+2211   &         16.5  \\
J164419.9+454644    &    2.27    &    1.16    &    44.21    &    1.3    &    0.38    &    1.82    &    0.2    &    20.1    &                     0.225    &                    &   1BIGB J164419.9+4546   &         16.3  \\
J165249.9+402309    &    7.05    &    2.46    &    44.36    &    2.74    &    0.63    &    2.0    &    0.18    &    29.49    &    >                0.310    &                    &   --   &         15.5  \\
J174702.5+493800    &    11.11    &    1.39    &    33.52    &    1.56    &    0.44    &    1.95    &    0.22    &    26.01    &                     0.460    &    T               &   1BIGB J174702.5+4938   &         17.0  \\
J184822.4+653656    &    9.77    &    0.89    &    47.14    &    0.94    &    0.34    &    1.56    &    0.18    &    20.02    &                     0.364    &                    &   1BIGB J184822.4+6536   &         17.7  \\
J194455.0-214318    &    74.63    &    9.18    &    524.94    &    10.28    &    0.91    &    1.81    &    0.07    &    75.71    &    >                0.410    &                    &   --   &         16.0  \\
J213852.6-205347    &    7.97    &    1.45    &    60.01    &    1.56    &    0.45    &    1.62    &    0.16    &    19.66    &                     0.290    &                    &   2FGL J2139.1-2054   &         17.0  \\
J224910.6-130002    &    19.48    &    4.01    &    73.38    &    4.02    &    0.66    &    2.39    &    0.2    &    140.67    &    >                0.500    &                    &   1BIGB J224910.6-1300   &         17.5  \\
J225147.5-320611    &    4.63    &    1.57    &    55.04    &    1.73    &    0.5    &    1.73    &    0.19    &    25.27    &                     0.246    &                    &   1BIGB J225147.5-3206   &         18.0  \\
J002200.0-514023    &    10.03    &    6.0    &    372.36    &    19.59    &    2.18    &    2.02    &    0.08    &    58.56    &                     0.250    &                    &   3FGL J0022.1-5141   &         15.7  \\
J003020.4-164712    &    8.95    &    3.93    &    187.25    &    1.72    &    0.22    &    1.81    &    0.11    &    58.11    &                     0.237    &                    &   3FGL J0030.2-1646   &         15.6  \\
J003334.3-192132    &    387.12    &    28.57    &    3214.1    &    47.87    &    2.04    &    1.8    &    0.03    &    105.89    &    >                0.506    &                    &   3FGL J0033.6-1921   &         15.7  \\
J004334.0-044300    &    20.48    &    1.71    &    45.21    &    0.48    &    0.12    &    1.8    &    0.2    &    26.95    &    >                0.480    &                    &   3FGL J0043.5-0444   &         16.7  \\
J004348.6-111606    &    5.91    &    2.44    &    68.34    &    0.73    &    0.14    &    1.9    &    0.15    &    26.28    &                     0.264    &                    &   3FGL J0043.7-1117   &         15.7  \\
J005116.6-624203    &    70.01    &    15.13    &    1847.48    &    10.64    &    0.54    &    1.73    &    0.04    &    127.56    &    >                0.300    &                    &   3FGL J0051.2-6241   &         15.9  \\
J011130.1+053626    &    5.91    &    1.16    &    28.78    &    0.85    &    0.27    &    1.84    &    0.22    &    26.15    &                     0.346    &                    &   3FGL J0111.5+0535   &         16.5  \\
J011904.6-145858    &    45.54    &    2.82    &    134.42    &    1.82    &    0.29    &    1.77    &    0.12    &    27.01    &    >                0.530    &                    &   3FGL J0118.9-1457   &         16.1  \\
J012338.2-231058    &    39.29    &    5.71    &    299.7    &    9.61    &    1.14    &    1.87    &    0.08    &    59.86    &                     0.404    &                    &   3FGL J0123.7-2312   &         17.3  \\
J015646.0-474417    &    4.49    &    1.99    &    61.6    &    1.5    &    0.3    &    2.04    &    0.19    &    31.64    &    >                0.290    &                    &   3FGL J0156.9-4742   &         16.6  \\
J020838.1+352312    &    10.47    &    2.42    &    84.95    &    0.69    &    0.12    &    1.82    &    0.15    &    24.63    &                     0.318    &                    &   3FGL J0208.6+3522   &         16.3  \\
J021252.7+224452    &    39.57    &    7.42    &    269.67    &    33.49    &    4.02    &    2.17    &    0.08    &    54.67    &                     0.459    &                    &   3FGL J0213.0+2245   &         15.2  \\
J021650.8-663642    &    21.68    &    7.57    &    465.79    &    16.52    &    1.4    &    2.08    &    0.08    &    66.16    &    >                0.330    &                    &   3FGL J0217.0-6635   &         15.5  \\
J022716.4+020159    &    42.23    &    5.95    &    266.31    &    16.94    &    2.12    &    1.98    &    0.08    &    29.43    &                     0.450    &                    &   3FGL J0227.2+0201   &         17.6  \\
J023734.0-360328    &    21.97    &    3.19    &    134.68    &    2.79    &    0.43    &    1.89    &    0.12    &    43.5    &    >                0.411    &                    &   3FGL J0237.5-3603   &         16.0  \\
J023832.3-311656    &    22.59    &    10.19    &    835.93    &    14.51    &    1.1    &    1.8    &    0.05    &    63.3    &                     0.233    &                    &   3FGL J0238.4-3117   &         16.3  \\
J024440.1-581954    &    18.05    &    5.29    &    411.39    &    1.82    &    0.17    &    1.72    &    0.08    &    88.32    &                     0.260    &                    &   3FGL J0244.8-5818   &         16.8  \\
J030326.3-240710    &    127.67    &    55.28    &    7499.57    &    340.74    &    12.35    &    1.93    &    0.02    &    672.76    &                     0.266    &                    &   3FGL J0303.4-2407   &         15.7  \\
J030416.3-283217    &    61.21    &    1.56    &    62.22    &    0.45    &    0.1    &    1.68    &    0.16    &    26.51    &    >                0.700    &                    &   3FGL J0304.3-2836   &         17.7  \\
J032523.5-563544    &    67.9    &    4.9    &    262.21    &    31.81    &    4.85    &    1.99    &    0.07    &    40.94    &                     0.600    &                    &   3FGL J0325.2-5634   &         16.5  \\
J032540.9-164615    &    35.67    &    10.75    &    754.22    &    10.94    &    0.8    &    1.85    &    0.06    &    67.4    &                     0.291    &                    &   3FGL J0325.6-1648   &         15.6  \\
J033812.4-244350    &    4.95    &    0.96    &    43.72    &    0.05    &    0.02    &    1.52    &    0.19    &    23.79    &                     0.251    &                    &   3FGL J0338.1-2443   &         17.1  \\
J041652.3+010522    &    23.01    &    6.7    &    317.69    &    2.76    &    0.28    &    1.82    &    0.08    &    36.76    &                     0.287    &                    &   3FGL J0416.8+0104   &         16.5  \\
J050534.6+041553    &    41.26    &    6.32    &    204.81    &    3.17    &    0.38    &    1.95    &    0.11    &    39.44    &                     0.424    &                    &   3FGL J0505.5+0416   &         15.7  \\
J050657.7-543503    &    25.07    &    5.05    &    488.15    &    2.21    &    0.21    &    1.56    &    0.07    &    40.57    &    >                0.260    &                    &   3FGL J0506.9-5435   &         16.2  \\
J050756.0+673723    &    265.31    &    17.08    &    2720.35    &    5.27    &    0.23    &    1.54    &    0.03    &    95.0    &                     0.416    &    T               &   3FGL J0508.0+6736   &         17.9  \\
J053628.9-334301    &    55.09    &    19.18    &    1336.69    &    379.34    &    33.18    &    2.12    &    0.03    &    156.11    &    >                0.340    &                    &   3FGL J0536.4-3347   &         16.0  \\
J054357.1-553207    &    54.18    &    15.6    &    1592.31    &    15.17    &    0.83    &    1.76    &    0.04    &    77.78    &                     0.273    &                    &   3FGL J0543.9-5531   &         16.7  \\
J055806.4-383830    &    21.14    &    6.14    &    347.52    &    8.0    &    0.84    &    1.87    &    0.08    &    42.16    &                     0.302    &                    &   3FGL J0558.1-3838   &         16.7  \\
J060408.5-481725    &    20.5    &    2.69    &    132.76    &    0.26    &    0.05    &    1.73    &    0.11    &    37.13    &    >                0.370    &                    &   3FGL J0604.1-4817   &         16.2  \\
J064443.6-285115    &    28.03    &    2.53    &    67.3    &    4.2    &    0.96    &    1.86    &    0.12    &    22.47    &    >                0.490    &                    &   3FGL J0644.6-2853   &         16.1  \\
J065046.3+250258    &    58.54    &    32.21    &    3411.1    &    25.7    &    0.96    &    1.75    &    0.03    &    188.73    &                     0.203    &    T               &   3FGL J0650.7+2503   &         16.8  \\
J074405.2+743357    &    21.21    &    4.58    &    372.66    &    7.65    &    0.87    &    1.78    &    0.07    &    47.18    &                     0.314    &                    &   3FGL J0744.3+7434   &         16.7  \\
J080457.7-062425    &    36.6    &    4.97    &    166.99    &    2.57    &    0.34    &    1.91    &    0.11    &    37.46    &    >                0.430    &                    &   3FGL J0805.0-0622   &         16.9  \\
J080625.9+593106    &    7.68    &    2.51    &    110.66    &    1.73    &    0.28    &    1.93    &    0.13    &    40.34    &                     0.300    &    T               &   3FGL J0806.6+5933   &         15.3  \\
J081627.1-131152    &    114.61    &    18.07    &    1336.78    &    17.85    &    1.01    &    1.81    &    0.04    &    114.91    &    >                0.370    &                    &   3FGL J0816.4-1311   &         16.4  \\
J082706.1-070844    &    9.54    &    4.04    &    155.31    &    7.63    &    1.25    &    1.84    &    0.09    &    36.62    &                     0.247    &    T               &   3FGL J0827.2-0711   &         16.3  \\
J090534.9+135805    &    48.19    &    9.5    &    651.36    &    13.0    &    1.07    &    1.82    &    0.06    &    44.29    &    >                0.340    &                    &   3FGL J0905.5+1358   &         15.1  \\
J091037.0+332924    &    58.77    &    14.17    &    1143.74    &    15.9    &    0.91    &    1.95    &    0.05    &    107.14    &                     0.350    &    T               &   3FGL J0910.5+3329   &         15.0  \\
J091230.5+155527    &    3.47    &    2.62    &    82.34    &    0.1    &    0.03    &    1.95    &    0.11    &    50.43    &                     0.212    &                    &   3FGL J0912.7+1556   &         16.9  \\
J091714.5-034314    &    9.99    &    1.44    &    48.02    &    0.62    &    0.17    &    1.59    &    0.17    &    28.02    &                     0.308    &                    &   3FGL J0917.3-0344   &         16.6  \\
J092542.7+595815    &    44.5    &    1.67    &    67.85    &    0.43    &    0.08    &    1.86    &    0.15    &    17.95    &    >                0.700    &                    &   3FGL J0925.6+5959   &         15.5  \\
J094022.3+614825    &    3.56    &    3.47    &    146.74    &    25.57    &    4.66    &    2.07    &    0.09    &    40.55    &                     0.210    &                    &   3FGL J0941.0+6151   &         16.2  \\
J094620.2+010450    &    34.71    &    1.71    &    44.65    &    0.33    &    0.08    &    1.76    &    0.19    &    21.72    &                     0.577    &                    &   3FGL J0946.2+0103   &         17.9  \\
J095805.8-031739    &    34.06    &    0.88    &    25.94    &    0.18    &    0.06    &    1.53    &    0.22    &    26.24    &    >                0.600    &                    &   3FGL J0958.3-0318   &         16.4  \\
J101244.2+422957    &    19.77    &    2.78    &    157.36    &    1.01    &    0.14    &    1.74    &    0.11    &    29.31    &                     0.365    &                    &   3FGL J1012.7+4229   &         16.8  \\
J102339.7+300056    &    12.89    &    1.55    &    49.01    &    0.51    &    0.12    &    1.86    &    0.19    &    27.66    &                     0.433    &                    &   3FGL J1023.7+3000   &         15.8  \\
J103118.4+505335    &    41.89    &    9.22    &    1014.96    &    7.26    &    0.48    &    1.74    &    0.05    &    52.72    &    >                0.360    &                    &   3FGL J1031.2+5053   &         17.0  \\
J104149.0+390118    &    2.05    &    1.85    &    53.25    &    2.71    &    0.59    &    2.03    &    0.17    &    48.69    &                     0.210    &                    &   3FGL J1041.8+3901   &         16.5  \\
J104651.4-253544    &    4.98    &    2.04    &    56.87    &    1.29    &    0.28    &    1.83    &    0.16    &    30.21    &                     0.250    &                    &   3FGL J1046.9-2531   &         18.0  \\
J105125.3+394324    &    30.66    &    2.58    &    101.98    &    1.19    &    0.19    &    1.84    &    0.14    &    32.63    &                     0.497    &                    &   3FGL J1051.4+3941   &         16.8  \\
J110124.7+410847    &    41.2    &    2.18    &    101.33    &    0.79    &    0.14    &    1.8    &    0.14    &    29.06    &    >                0.580    &                    &   3FGL J1101.5+4106   &         15.7  \\
J110747.9+150209    &    11.18    &    6.18    &    305.15    &    14.09    &    1.6    &    1.98    &    0.08    &    37.98    &                     0.250    &    T               &   3FGL J1107.8+1502   &         15.6  \\
J111224.5+175120    &    8.97    &    1.14    &    25.7    &    0.07    &    0.03    &    1.85    &    0.22    &    25.04    &                     0.420    &                    &   3FGL J1112.6+1749   &         16.9  \\
J111939.4-304720    &    13.95    &    0.62    &    28.6    &    0.1    &    0.04    &    1.38    &    0.23    &    23.38    &                     0.412    &                    &   3FGL J1119.7-3046   &         17.1  \\
J112453.8+493409    &    93.35    &    4.89    &    362.65    &    2.91    &    0.29    &    1.78    &    0.08    &    63.95    &    >                0.570    &                    &   3FGL J1124.9+4932   &         16.5  \\
J112551.9-074220    &    6.16    &    1.86    &    47.16    &    1.78    &    0.47    &    1.81    &    0.16    &    22.16    &                     0.279    &                    &   3FGL J1125.8-0745   &         15.7  \\
J114930.3+243925    &    9.84    &    1.35    &    31.19    &    1.28    &    0.43    &    1.84    &    0.19    &    23.45    &                     0.402    &                    &   3FGL J1149.5+2443   &         17.1  \\
J115034.6+415439    &    79.88    &    19.26    &    2122.31    &    29.82    &    1.46    &    1.85    &    0.04    &    85.54    &    >                0.320    &                    &   3FGL J1150.5+4155   &         15.6  \\
J115404.5-001009    &    13.36    &    3.99    &    202.3    &    1.47    &    0.2    &    1.71    &    0.1    &    46.77    &                     0.254    &                    &   3FGL J1154.2-0010   &         16.6  \\
J115853.2+081942    &    5.86    &    1.84    &    50.53    &    1.12    &    0.25    &    1.87    &    0.19    &    25.99    &                     0.290    &                    &   3FGL J1158.9+0818   &         16.1  \\
J120412.1+114555    &    9.78    &    3.91    &    125.3    &    3.26    &    0.47    &    2.01    &    0.13    &    46.81    &                     0.296    &                    &   3FGL J1204.0+1144   &         16.6  \\
J121945.7-031422    &    15.32    &    5.15    &    200.41    &    9.2    &    1.3    &    1.94    &    0.09    &    51.71    &                     0.299    &                    &   3FGL J1219.7-0314   &         16.0  \\
J122424.1+243623    &    20.61    &    14.18    &    1081.89    &    16.98    &    1.02    &    1.93    &    0.05    &    226.87    &                     0.218    &                    &   3FGL J1224.5+2436   &         16.1  \\
J122644.2+063853    &    55.02    &    2.04    &    94.42    &    0.37    &    0.07    &    1.65    &    0.14    &    22.51    &                     0.583    &                    &   3FGL J1226.8+0638   &         15.8  \\
J123123.8+142124    &    10.65    &    3.27    &    128.07    &    4.74    &    0.87    &    1.73    &    0.09    &    22.81    &                     0.256    &                    &   3FGL J1231.8+1421   &         15.4  \\
J123738.9+625841    &    3.23    &    0.8    &    27.79    &    0.42    &    0.13    &    1.78    &    0.22    &    29.47    &                     0.297    &                    &   3FGL J1237.9+6258   &         16.0  \\
J124312.7+362743    &    99.06    &    21.98    &    2594.81    &    33.1    &    1.52    &    1.78    &    0.03    &    62.08    &    >                1.065   &                    &   3FGL J1243.1+3627   &         16.2  \\
J131532.5+113330    &    44.19    &    2.44    &    63.8    &    1.83    &    0.36    &    1.88    &    0.15    &    39.18    &    >                0.610    &                    &   3FGL J1315.4+1130   &         16.7  \\
J132301.0+043951    &    2.69    &    2.22    &    49.54    &    0.76    &    0.16    &    2.06    &    0.19    &    22.21    &                     0.224    &                    &   3FGL J1322.9+0435   &         16.8  \\
J132358.3+140558    &    53.68    &    5.75    &    271.93    &    13.23    &    1.65    &    1.89    &    0.08    &    36.22    &    >                0.470    &                    &   3FGL J1323.9+1405   &         15.4  \\
J134029.8+441004    &    39.93    &    1.62    &    74.83    &    0.7    &    0.16    &    1.61    &    0.14    &    36.4    &                     0.540    &                    &   3FGL J1340.6+4412   &         17.3  \\
J140450.8+040202    &    25.64    &    4.39    &    183.09    &    7.71    &    1.14    &    1.85    &    0.09    &    24.03    &    >                0.370    &                    &   3FGL J1404.8+0401   &         15.7  \\
J140659.1+164206    &    26.99    &    1.46    &    44.26    &    0.33    &    0.08    &    1.73    &    0.18    &    24.55    &    >                0.540    &                    &   3FGL J1406.6+1644   &         17.0  \\
J141756.5+254324    &    10.14    &    2.86    &    147.3    &    7.2    &    1.65    &    1.63    &    0.08    &    24.58    &                     0.237    &                    &   3FGL J1417.8+2540   &         17.4  \\
J141826.2-023333    &    127.76    &    28.2    &    2765.72    &    57.51    &    2.33    &    1.48    &    nan    &    134.68    &    >                0.356    &                    &   3FGL J1418.4-0233   &         15.5  \\
J141900.3+773229    &    12.81    &    4.34    &    328.47    &    5.14    &    0.53    &    1.83    &    0.07    &    32.38    &    >                0.270    &                    &   3FGL J1418.9+7731   &         16.0  \\
J143657.7+563924    &    57.3    &    5.79    &    522.39    &    8.28    &    0.8    &    1.77    &    0.06    &    47.4    &    >                0.430    &                    &   3FGL J1436.8+5639   &         16.9  \\
J143917.3+393242    &    18.93    &    5.28    &    289.04    &    3.68    &    0.37    &    2.01    &    0.1    &    42.05    &                     0.344    &                    &   3FGL J1439.2+3931   &         15.9  \\
J144037.7-384654    &    40.0    &    5.42    &    285.18    &    1.5    &    0.16    &    1.68    &    0.08    &    36.35    &    >                0.350    &                    &   3FGL J1440.4-3845   &         17.2  \\
J144506.1-032612    &    27.6    &    6.55    &    281.39    &    10.13    &    1.24    &    1.81    &    0.07    &    35.43    &    >                0.310    &                    &   3FGL J1445.0-0328   &         17.4  \\
J145127.7+635419    &    52.03    &    1.61    &    96.65    &    1.03    &    0.2    &    1.68    &    0.13    &    29.58    &                     0.650    &                    &   3FGL J1451.2+6355   &         17.0  \\
J150101.7+223806    &    17.13    &    11.95    &    746.59    &    25.58    &    1.81    &    2.03    &    0.06    &    106.44    &                     0.235    &                    &   3FGL J1500.9+2238   &         15.1  \\
J150340.6-154113    &    65.09    &    9.18    &    439.69    &    5.04    &    0.42    &    1.79    &    0.07    &    52.0    &    >                0.380    &                    &   3FGL J1503.7-1540   &         17.6  \\
J150716.3+172102    &    56.74    &    3.44    &    131.85    &    1.5    &    0.21    &    1.83    &    0.11    &    43.84    &                     0.565    &                    &   3FGL J1507.4+1725   &         15.7  \\
J150842.5+270908    &    6.95    &    1.54    &    58.39    &    0.48    &    0.11    &    1.64    &    0.16    &    23.28    &                     0.270    &                    &   3FGL J1508.6+2709   &         17.8  \\
J153311.2+185429    &    12.93    &    2.56    &    117.52    &    1.86    &    0.35    &    1.71    &    0.11    &    28.7    &                     0.305    &                    &   3FGL J1533.2+1852   &         17.2  \\
J153500.7+532036    &    39.18    &    1.48    &    68.96    &    1.78    &    0.44    &    1.67    &    0.12    &    32.12    &    >                0.590    &                    &   3FGL J1534.4+5323   &         17.2  \\
J154604.2+081913    &    28.91    &    6.47    &    318.97    &    2.55    &    0.25    &    1.91    &    0.08    &    65.78    &    >                0.350    &                    &   3FGL J1546.0+0818   &         15.1  \\
J154712.1-280221    &    53.33    &    3.12    &    80.1    &    0.93    &    0.16    &    1.83    &    0.14    &    30.87    &    >                0.570    &                    &   3FGL J1547.1-2801   &         15.8  \\
J155424.1+201125    &    5.85    &    2.15    &    64.0    &    6.92    &    1.95    &    1.88    &    0.11    &    26.73    &                     0.273    &                    &   3FGL J1554.4+2010   &         17.4  \\
J155543.0+111123    &    878.53    &    140.08    &    26507.02    &    468.11    &    7.41    &    1.42    &    nan    &    224.09    &    >                0.443    &                    &   3FGL J1555.7+1111   &         15.6  \\
J160620.8+563016    &    15.78    &    1.48    &    63.57    &    0.88    &    0.2    &    1.78    &    0.16    &    32.23    &                     0.450    &                    &   3FGL J1606.1+5630   &         16.0  \\
J162625.8+351341    &    21.79    &    1.64    &    61.92    &    0.48    &    0.1    &    1.79    &    0.16    &    29.82    &                     0.498    &                    &   3FGL J1626.1+3512   &         16.0  \\
J170238.5+311542    &    39.15    &    3.54    &    180.58    &    0.51    &    0.07    &    1.81    &    0.09    &    37.09    &    >                0.470    &                    &   3FGL J1702.6+3116   &         15.4  \\
J175615.9+552218    &    45.88    &    3.68    &    223.86    &    3.85    &    0.5    &    1.76    &    0.08    &    32.04    &    >                0.470    &                    &   3FGL J1756.3+5523   &         17.3  \\
J175713.0+703337    &    17.18    &    2.45    &    90.8    &    0.31    &    0.06    &    1.87    &    0.13    &    34.16    &                     0.407    &                    &   3FGL J1756.9+7032   &         17.3  \\
J183849.0+480234    &    81.84    &    20.04    &    2206.35    &    20.22    &    0.93    &    1.79    &    0.04    &    216.67    &                     0.300    &    T               &   3FGL J1838.8+4802   &         15.8  \\
J201428.6-004721    &    7.47    &    4.92    &    133.66    &    3.21    &    0.44    &    1.98    &    0.12    &    48.3    &                     0.231    &    T               &   3FGL J2014.3-0047   &         15.2  \\
J201624.0-090333    &    49.24    &    11.68    &    589.46    &    27.26    &    2.22    &    2.0    &    0.06    &    51.79    &                     0.367    &    T               &   3FGL J2016.4-0905   &         15.0  \\
J203649.3-332830    &    7.66    &    2.31    &    77.87    &    0.12    &    0.03    &    1.62    &    0.12    &    24.53    &                     0.230    &                    &   3FGL J2036.6-3325   &         16.3  \\
J205528.2-002116    &    32.19    &    2.96    &    92.1    &    0.8    &    0.14    &    1.75    &    0.13    &    27.67    &                     0.440    &                    &   3FGL J2055.2-0019   &         18.0  \\
J213103.1-274656    &    49.59    &    8.82    &    543.82    &    4.85    &    0.38    &    1.9    &    0.07    &    49.09    &    >                0.380    &                    &   3FGL J2130.8-2745   &         16.1  \\
J213135.3-091523    &    56.54    &    6.51    &    291.27    &    14.19    &    1.76    &    1.88    &    0.07    &    56.71    &                     0.449    &                    &   3FGL J2131.5-0915   &         16.4  \\
J213151.4-251557    &    151.09    &    3.49    &    136.81    &    5.01    &    0.82    &    1.86    &    0.11    &    24.51    &    >                0.860    &                    &   3FGL J2131.8-2516   &         16.9  \\
J214552.1+071927    &    3.13    &    2.75    &    61.8    &    89.18    &    37.36    &    2.18    &    0.09    &    31.31    &                     0.237    &                    &   3FGL J2145.7+0717   &         17.5  \\
J214636.9-134359    &    120.94    &    12.42    &    901.51    &    6.88    &    0.46    &    1.75    &    0.05    &    72.84    &    >                0.420    &                    &   3FGL J2146.6-1344   &         15.7  \\
J215015.4-141049    &    8.47    &    3.92    &    157.35    &    2.51    &    0.36    &    1.76    &    0.1    &    28.71    &                     0.220    &                    &   3FGL J2150.2-1411   &         17.8  \\
J215305.2-004229    &    11.43    &    2.66    &    61.44    &    2.04    &    0.47    &    1.9    &    0.16    &    35.95    &                     0.341    &                    &   3FGL J2152.9-0045   &         18.0  \\
J222129.2-522527    &    39.59    &    8.71    &    634.52    &    16.4    &    1.36    &    1.87    &    0.06    &    66.99    &    >                0.340    &                    &   3FGL J2221.6-5225   &         15.8  \\
J225818.9-552536    &    19.73    &    2.97    &    123.74    &    19.87    &    4.03    &    2.1    &    0.1    &    38.22    &                     0.479    &                    &   3FGL J2258.3-5526   &         15.7  \\
J230722.0-120517    &    15.01    &    1.13    &    30.63    &    0.57    &    0.19    &    1.73    &    0.21    &    23.05    &    >                0.470    &                    &   3FGL J2307.4-1208   &         16.5  \\
J232444.5-404049    &    27.82    &    10.72    &    866.67    &    7.26    &    0.51    &    1.76    &    0.06    &    84.18    &    >                0.240    &                    &   3FGL J2324.7-4040   &         15.5  \\
J235034.3-300603    &    6.06    &    3.93    &    153.56    &    8.61    &    1.3    &    1.96    &    0.11    &    51.19    &                     0.230    &                    &   3FGL J2350.4-3004   &         15.7  \\
J235612.1+403643    &    14.84    &    4.02    &    147.69    &    1.19    &    0.16    &    1.95    &    0.11    &    41.97    &                     0.331    &                    &   3FGL J2356.0+4037   &         16.3  \\
\end{longtable}

% Don't change these lines
\bsp	% typesetting comment
\label{lastpage}
\end{document}